\documentclass[aps,prd,twocolumn,superscriptaddress,showpacs,showkeys]{revtex4}

\usepackage{epsfig,epsf}
\usepackage{amsmath}
\usepackage{amsthm}
\usepackage{amsfonts}
\usepackage{amssymb}
\usepackage{dsfont}

\usepackage{multirow}

\usepackage{slashed}

\usepackage[active]{srcltx}
\usepackage{wick}
\newcommand{\be}{\begin{equation}}
\newcommand{\ee}{\end{equation}}
\newcommand{\ba}{\begin{eqnarray}}
\newcommand{\ea}{\end{eqnarray}}
\newcommand{\baa}{\begin{eqnarray*}}
\newcommand{\btab}{\begin{tabular}}
\newcommand{\etab}{\end{tabular}}
\newcommand{\eaa}{\end{eqnarray*}}

\newcommand \vev [1] {\langle{#1}\rangle}

\def\inbar{\,\vrule height1.5ex width.4pt depth0pt}
\def\IC{\relax\hbox{$\inbar\kern-.3em{\rm C}$}}
\def\IZ{\relax{\hbox{\cmss Z\kern-.4em Z}}}
\def\IR{{\hbox{{\rm I}\kern-.2em\hbox{\rm R}}}}

\def\IP{{\hbox{{\rm I}\kern-.2em\hbox{\rm P}}}}
\def\II{\hbox{{1}\kern-.25em\hbox{l}}}

\begin{document}

\title{ Scale dependence of twist-three contributions to single
     spin asymmetries}


\date{\today}

\author{ V.M.~Braun}
\affiliation{Institut f\"ur Theoretische Physik, Universit\"at
   Regensburg,D-93040 Regensburg, Germany}
\author{ A.N.~Manashov}
\affiliation{Institut f\"ur Theoretische Physik, Universit\"at
   Regensburg,D-93040 Regensburg, Germany}
\affiliation{Department of Theoretical Physics,  St.-Petersburg State
University\\
199034, St.-Petersburg, Russia}
\author{B.~Pirnay}
\affiliation{Institut f\"ur Theoretische Physik, Universit\"at
   Regensburg,D-93040 Regensburg, Germany}

\date{\today}

\begin{abstract}
We reexamine the scale dependence of twist-three correlation functions
relevant for the single transverse spin asymmetry in the framework
of collinear factorization.
Evolution equations are derived for both the flavor--nonsinglet
and flavor--singlet distributions and arbitrary parton momenta.
Our results do not agree with the recent calculations of the evolution in
the limit of vanishing gluon momentum. Possible sources for this discrepancy
are identified.

\end{abstract}



\maketitle


%
\section{Introduction}
%
Large transverse single spin asymmetries (SSAs) have been observed in different hadronic
reactions and these observations generated a lot of interest. Such experiments are
conceptually rather simple, but their theoretical description proved to be challenging
as the leading-twist contributions to such asymmetries
vanish, see \cite{Anselmino:1994gn,Liang:2000gz,Barone:2001sp} for a review.
Over the past few years there was a splash of theoretical activity in this field,
which mainly followed two lines: the $k_\perp$ factorization in terms of
the transverse-momentum dependent (TMD) distributions (e.g.
\cite{Sivers:1990fh,Mulders:1995dh,Boer:1997bw,Brodsky:2002cx,Brodsky:2002rv,Collins:2002kn,Boer:2003cm,Bacchetta:2006tn}),
or, alternatively, collinear factorization including twist-three contributions in terms
of multiparton correlation functions
\cite{Efremov:1981sh,Efremov:1984ip,Qiu:1991pp,Qiu:1991wg,Efremov:1994dg,Qiu:1998ia,Kanazawa:2000hz,Kanazawa:2000cx,Eguchi:2006mc,Koike:2007rq,Kang:2008ih}.
 These two techniques have their own domain of validity and
were shown to be consistent with each other in the kinematic regime where they both apply
\cite{Ji:2006ub,Ji:2006vf,Ji:2006br,Koike:2007dg}.

However, practically all existing calculations have been so far at the leading order (LO) which
corresponds, roughly speaking, to the (generalized) parton model. In order to test the QCD
dynamics and eventually also reduce the dependence of theory predictions on both the factorization
scale and the renormalization scale of the strong coupling, it is necessary to calculate
the scale dependence (evolution) of the relevant nonperturbative functions. The
corresponding  calculation was done in Refs.~\cite{Kang:2008ey,Vogelsang:2009pj} where
the evolution equation was derived for the gluonic pole contributions to
twist-three correlation functions that are relevant to SSA. Unfortunately,
it turns out that the equations derived in \cite{Kang:2008ey,Vogelsang:2009pj}
are in contradiction to earlier results obtained in the framework of the operator
product expansion (OPE). A resolution of this discrepancy presents the main motivation
for this study.

Although we do reproduce the terms that appear in Ref.~\cite{Kang:2008ey}, we obtain two
additional contributions to the evolution  equation both for flavor-nonsinget and
flavor-singlet operators. We have found that part of the disagreement is due to omission
of cut diagrams corresponding to the mixing of conventional twist-three correlation
functions that are discussed in the context of SSA with the contributions involving a
gluon and the quark-antiquark pair on the opposite sides of the cut. These are the same
contributions that have been discussed for some time for semi-inclusive
DIS~\cite{Eguchi:2006mc,Koike:2007dg,Koike:2009yb},
and the new observation in this work is that they necessarily contribute to the leading
logarithmic accuracy through the (factorization) scale dependence. For another part we do
not have an intuitive explanation. Our OPE-based calculation contains additional
contributions (compared to  \cite{Kang:2008ey,Vogelsang:2009pj}) for which sending the
gluon momentum to zero has to be done with caution, as will be explained in the text.

Another question that we want to address is to clarify the relation of these results
to earlier calculations of the scale dependence of twist-three correlation functions
that contribute to inclusive reactions, pioneered by the study of the structure
function $g_2(x,Q^2)$ in Ref.~\cite{Bukhvostov:1984as}.
We would like to emphasize that the scale dependence of an {\it arbitrary} twist-three operator
in QCD, hence {\it arbitrary} twist-three light-cone correlation function can be
determined in terms of the two-particle evolution kernels introduced by
Bukhvostov, Frolov, Lipatov and Kuraev (BFLK) \cite{Bukhvostov:1985rn}.
The BFLK approach has become standard in calculations of the spectrum of anomalous
dimensions (dilatation operator) in supersymmetric theories that are relevant to the
AdS/CFT correspondence (see e.g. \cite{Belitsky:2004cz,Beisert:2004ry}),
but, unfortunately, remains to be largely unknown to the broad
QCD community. We will show that evolution equations for the particular
parton distributions relevant for SSA \cite{Kang:2008ey} do not require an independent
calculation but can be obtained from the BFLK kernels by simple algebra.
Using an updated version of this technique \cite{Braun:2008ia,Braun:2009vc}
we derive the complete evolution equations for all relevant three-particle correlation functions
for arbitrary gluon momenta. This generalization is necessary since the gluon-pole
contributions considered in \cite{Kang:2008ey,Vogelsang:2009pj} do not have autonomous
scale dependence.

The presentation is organized as follows. Section 2 is introductory. We review basic
properties of three-particle correlation functions and introduce a very convenient
decomposition of momenta (coordinates) and the field operators in the spinor
representation. This rewriting makes the symmetries explicit and drastically simplifies
the forthcoming algebra. Section 3 contains a detailed derivation of the flavor-nonsinglet
evolution equation in the BFLK approach and the comparison with
\cite{Kang:2008ey,Vogelsang:2009pj}. Our result coincides identically with the
corresponding equation for the quark-antiquark-gluon correlation function relevant for the
structure function $g_2(x,Q^2)$. The flavor-singlet evolution is considered in Section 4.
In this case there are two independent evolution equations for positive and negative
C-parity that involve three-gluon correlation functions involving $SU(3)$ structure
constants $f^{abc}$ and  $d^{abc}$, respectively. The equation for positive C-parity
coincides, again, with the corresponding equation for the structure function $g_2(x,Q^2)$,
whereas the equation for negative C-parity is, to our knowledge, a new result (in this
form). Finally, in Section 5 we summarize. The full list of the BFLK kernels in the
momentum representation is collected in the Appendix.

%
\section{General discussion}
%

\subsection{Definition and support properties of three-particle light-cone correlation functions}

{}Following Ref.~\cite{Kang:2008ey} we will consider
the correlation functions $\widetilde{\mathcal{T}}_{q,F}$,
$\widetilde{\mathcal{T}}_{\Delta q,F}$, corresponding to the
nucleon matrix elements  of the quark-antiquark-gluon light-ray operators
\begin{align}\label{TdT}
T_\mu(z_1,z_2,z_3)&=g\bar q(z_1 n)\gamma_+ F_{\mu+}(z_2 n) q(z_3 n)\,,
\nonumber\\
\Delta T_\mu(z_1,z_2,z_3)&=g\bar q(z_1 n)\gamma_+\gamma_5 iF_{\mu+}(z_2 n) q(z_3 n)\,
\end{align}
and also the correlation functions
$\widetilde{\mathcal{T}}^{(f,d)}_{G,F}$ and $\widetilde{\mathcal{T}}^{(f,d)}_{\Delta G,F}$
corresponding to the three-gluon operators
\begin{align}\label{FFF}
G_{\mu\rho\lambda}^{\pm}(z_1,z_2,z_3)&=
g\,C^{abc}_{\pm} F_{+\rho}^{a}(z_1 n) F_{+\mu}^b(z_2 n)
F_{+\lambda}^c(z_3 n)\,.
\end{align}
Here $n_\mu$ is the light-like vector, $n^2=0$, the ``plus'' projection
is defined as $a_+=a^\mu n_\mu$. Note that there are two gluon operators with
a different color structure; the factors $C^{abc}_{\pm}$ are written in terms
of the $SU(3)$ structure constants
\begin{align}\label{Cpm}
C^{abc}_+&=i\mathrm{f}^{abc}\,,
\notag\\
C^{abc}_-&=\mathrm{d}^{abc}\,.
\end{align}
In all cases the path-ordered Wilson lines are implied that ensure gauge
invariance. They are not shown for brevity.

\begin{widetext}
The definitions of parton distributions used by Kang and Qiu are
(cf. Eqs.~(12),(16),(14),(23) in~\cite{Kang:2008ey}):
\begin{align}\label{Qiu1}
\widetilde{\mathcal{T}}_{q,F}(x,x+x_2)&=\frac12\int \frac{dz_1 dz_2}{(2\pi)^2}e^{iP_+
(z_1 x+z_2 x_2)}
\vev{P,s_T|\tilde{s}^\mu T_\mu(0,z_2,z_1)|P,s_T}\,,
\notag\\
\widetilde{\mathcal{T}}_{\Delta q,F}(x,x+x_2)&=\frac12\int \frac{dz_2 dz_3}{(2\pi)^2}e^{iP_+
(z_1 x+z_2 x_2)}
\vev{P,s_T|{s}^\mu \Delta T_\mu(0,z_2,z_1)|P,s_T}\,,
\notag\\
\widetilde{\mathcal{T}}^{(f,d)}_{G,F}(x,x+x_2)&=\frac{g^{\rho\lambda}}{P^+}
\int \frac{dz_1 dz_2}{(2\pi)^2}e^{iP_+
(z_1 x+z_2 x_2)}
\vev{P,s_T|\tilde{s}^\mu G^\pm_{\mu\rho\lambda}(0,z_2,z_1)|P,s_T}\,,
\notag\\
\widetilde{\mathcal{T}}^{(f,d)}_{\Delta G,F}(x,x+x_2)&=\frac{\epsilon_{\perp}^{\rho\lambda}}{P^+}
\int \frac{dz_1 dz_2}{(2\pi)^2}e^{iP_+
(z_1 x+z_2 x_2)}
\vev{P,s_T|{s}^\mu G^\pm_{\mu\rho\lambda}(0,z_2,z_1)|P,s_T}\,.
\end{align}
Here $s_\mu$ is the nucleon spin vector normalized by the condition $s^2=-1$,
$\tilde s^\mu= -\epsilon^{\mu\nu\rho\sigma}s_\nu n_\rho \tilde n_\sigma$ with $\tilde n$
being the second  light-like vector, $\tilde n^2=0$, $n\tilde n=1$ and
$\epsilon_{\perp}^{\rho\lambda}=-\epsilon^{\rho\lambda n\tilde n}$,  $\epsilon_{0123}=1$.
The spin vector $s_\mu$  is assumed to be transverse, $ns=\tilde n s=0$.
In the last two equations $\widetilde{\mathcal{T}}^{(f)}$ and $\widetilde{\mathcal{T}}^{(d)}$
correspond to matrix elements of $G^+_{\mu\rho\lambda}$ and  $G^-_{\mu\rho\lambda}$, respectively, cf.
Eq.~(\ref{Cpm}).

{}For our purposes it is convenient to use a more symmetric notation with quark, antiquark
and gluon momentum fractions treated equally
\begin{align}\label{TDT}
\vev{P,s_T|\tilde{s}^\mu T_\mu(z_1,z_2,z_3)|P,s_T}
&=2P_+^2  \int \mathcal{D}x
\,e^{-iP_+(\sum_{k}x_k z_k)}\, T_{\bar q F q}(x_1,x_2,x_3)\,,
\notag\\
\vev{P,s_T|{s}^\mu \Delta T_\mu(z_1,z_2,z_3)|P,s_T}
&=2P_+^2  \int \mathcal{D}x
\,e^{-iP_+(\sum_{k}x_k z_k)}\, \Delta T_{\bar q F q}(x_1,x_2,x_3)\,,
\notag\\
g^{\rho\lambda}\vev{P,s_T|\tilde{s}^\mu G^{\pm}_{\mu\rho\lambda}(z_1,z_2,z_3)|P,s_T}&=
P_+^3  \int \mathcal{D}x
\,e^{-iP_+(\sum_{k}x_k z_k)}\, T^{\pm}_{3F}(x_1,x_2,x_3)\,,
\notag\\
\epsilon_{\perp}^{\rho\lambda}\vev{P,s_T|\tilde{s}^\mu G^{\pm}_{\mu\rho\lambda}(z_1,z_2,z_3)|P,s_T}&=
P_+^3 \int \mathcal{D}x
\,e^{-iP_+(\sum_{k}x_k z_k)}\, \Delta T^{\pm}_{3F}(x_1,x_2,x_3)\,,
\end{align}
\end{widetext}
where the integration measure is defined as
\begin{align}
 \int \mathcal{D}x = \int_{-1}^1 dx_1dx_2dx_3\,\delta(x_1+x_2+x_3)\,.
\label{measure}
\end{align}
We assume here the standard (relativistic) normalization of states; the factor $P_+^2$ is
necessary to ensure the reparametrization invariance to the choice of the light-cone
vector $n_\mu \to \alpha n_\mu$.
Obviously
\begin{align}
\widetilde{\mathcal{T}}_{q,F}(x,x+x_2)&\equiv  T_{\bar q F q}(-x-x_2,x_2,x)\,,
\end{align}
and similarly for the other distributions.
Writing the definition in such a form (Fourier-transformed)
makes explicit the support properties of the correlation functions: they are defined on the
surface $x_1+x_2+x_3=0$ and effectively are functions of two variables only.

\begin{figure}[t]
\includegraphics[width=5cm]{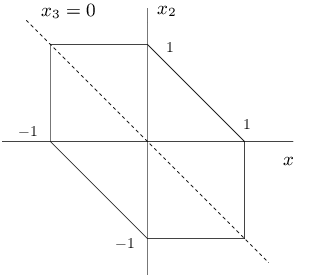}
\caption{Support properties of the parton correlation functions (\ref{Qiu1}) in the $(x,x_2)$ plane.
\label{figure1} }
\end{figure}

The support of parton correlation functions in the notation by Kang and Qiu,
i.e. in the $(x,x_2)$ plane, is shown in
Fig.~\ref{figure1}.
It can be separated in six different regions where $x$, $x_2$ and
$x+x_2$ are positive or negative, respectively. The parton-model interpretation of
each region is different. As discussed in detail in
Ref.~\cite{Jaffe:1983hp}, interpretation of light-cone correlation
functions as describing sequential emission or absorption of quark-, antiquark- and gluon-partons
by the target arises by choice of a particular representation in terms of the sum
over intermediate states (cut diagrams) that does not involve semidisconnected contributions.
In particular, the upper-right region in Fig.~\ref{figure1} corresponds to
emission of a pair of a quark-parton and a gluon with momentum fractions $x>0$, $x_2>0$,
respectively, and subsequent absorption of the quark-parton with $x+x_2$.

\begin{figure}[t]
\includegraphics[width=8cm]{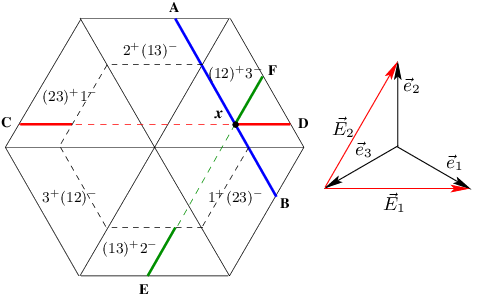}
\caption{Support properties of the correlation functions
(\ref{TDT}) in barycentric coordinates.
For the explanation of different regions and lines see text.
\label{figure2}}
\end{figure}

The picture can be made more symmetric going over to the correlation functions
(\ref{TdT}) and using analogue to barycentric coordinates~\cite{Mobius} as shown in
Fig.~\ref{figure2}:
\begin{align*}\label{bary}
\vec{x}=x_1\vec{e}_1+x_2\vec{e}_2 + x_3\vec{e}_3 = x_1\vec{E}_1+x_2\vec{E}_2\,.
\end{align*}
The six different regions, labeled $(12)^+3^-$, $2^+(13)^-$, etc.,
correspond to different subprocesses at the parton level~\cite{Jaffe:1983hp};
For each parton $k=1,2,3$\, ``plus'' stands for emission ($x_k>0$)
 and ``minus'' for absorption ($x_k<0$). Alternatively, one may think of
 ``plus'' and ``minus'' labels as indicating whether the corresponding parton
appears in the direct or the final amplitude in the cut diagram~\cite{Balitsky:1990ck}.

It is important that different regions do not have autonomous scale dependence;
they ``talk'' to each other and get mixed under the evolution. The particular mixing
pattern can be understood as follows.

It is easy to see that to one loop accuracy any contribution to the evolution equation
only involves two of the three partons. For example,  parton--1 and parton--2 can exchange
a gluon whereas parton--3 stays a spectator, etc. The full evolution kernel can, therefore,
be split in three two-particle kernels that involve parton pairs (12), (23) and (31),
respectively. Schematically
\begin{equation}
 \mu \frac{d}{\mu}  T_{\bar q F q} = \frac{\alpha_s}{2\pi}
[H_{12}+H_{23}+H_{31}]\otimes T_{\bar q F q}.
\label{111}
\end{equation}
Each two-particle kernel is, obviously, a function of the contributing parton momentum
fractions, e.g. $H_{12} \equiv H_{12}(x_1,x_2; x'_1,x'_2)$ and due to energy conservation
$x'_1+x'_2 = x_1+x_2$. In other words, the rate for the scale variation of
a three-particle parton correlation function
with given values of momentum fractions $x_1,x_2$ can only involve this function
on the line of constant  $x_1+x_2 = -x_3$ (for the contribution of $H_{12}$),
as shown in Fig.~\ref{figure2} (thick blue line {\small\it\bf AB}).

Since $x_3\not=0$ (in general), the total momentum fraction carried by the two
participating partons is non-zero, $x_1+x_2\not=0$. This situation
is familiar from studies of the scale dependence of leading-twist generalized parton
distributions (GPDs)
(see e.g. \cite{Diehl:2003ny,Belitsky:2005qn}) and the three regions $2^+(13)^-$,
$(12)^+3^-$, $1^+(23)^-$ traversed by the thick blue line {\small\bf AB}
in Fig.~\ref{figure2}
are in one-to-one correspondence to the two DGLAP regions ($2^+ 1^-$ and $1^+ 2^-$)
and the ERBL (central) region ($1^+2^+$) in the corresponding evolution equations. As well known
\cite{Diehl:2003ny,Belitsky:2005qn}, the scale dependence of GPDs in the DGLAP regions
is autonomous, whereas in the ERBL mode there are also terms describing ``leakage'' from
the DGLAP regions. In the present context, this result implies that
evolution equation for a generic three-particle light-cone correlation function
for the momentum fractions $x_1,x_2$ in the $(12)^+3^-$ region, as in  Fig.~\ref{figure2},
will receive nontrivial contributions from the $2^+(13)^-$, $1^+(23)^-$, $2^-(13)^+$ and
 $1^-(23)^+$ regions as well,
which have a different partonic interpretation.

So far we have considered the contribution of $H_{12}$ only.
The contributions of $H_{23}$ and $H_{13}$ are, in turn,  kinematically constrained
to the lines of constant $x'_2+x'_3$ and $x'_1+x'_3$, respectively, and, for the
particular choice of $x_1,x_2$ in the $(12)^+3^-$ region, correspond to the DGLAP modes
of the corresponding GPD-like evolution equations. Recall that the DGLAP evolution
is ordered in the momentum fraction; hence only parts of the kinematically
allowed regions contribute, as shown in Fig.~\ref{figure2} by the thick red
{\small\bf CD} ($H_{13}$)
and thick green {\small\bf EF} ($H_{23}$) lines.

Note that evolution of the parton correlation function for given values of $x_1,x_2$
only involves the regions of momentum fractions outwards from the center.
If one draws a small(er) hexagon (see Fig.~\ref{figure2}) inside the
big one, then evolution of the correlation function outside of the small hexagon does not
depend on the latter inside the small hexagon.
This ``radial'' ordering is a generalization of the usual momentum fraction
ordering in DGLAP equations. In contrast, there is no ordering/restrictions
in the ``azimuthal'' direction and, in principle, all regions ``talk'' to
each other. This property may result in the increase of the evolution rate
in the  small $x$ region.

\subsection{Spinor representation and symmetry properties}

Analysis of symmetry properties and the scale dependence of the light-cone correlation
functions can be simplified significantly by going over to the spinor representation.
Each covariant four-vector $x_\mu$ is mapped to a hermitian $2\times2$ matrix $x$:
 $$
x_{\alpha\dot\alpha}=x_\mu (\sigma^{\mu})_{\alpha\dot\alpha}\,,
\qquad
\bar x^{\dot\alpha\alpha}=x_\mu (\bar\sigma^{\mu})^{\dot\alpha\alpha}\,,
$$
where $\sigma^\mu=(\II,\vec{\sigma})$, $\bar\sigma^\mu=(\II,-\vec{\sigma})$ and
$\vec{\sigma}$ are the usual Pauli matrices. In components
\begin{align}\label{xd}
x=\begin{pmatrix}
x_0+x_3 & x_1-ix_2\\
x_1+ix_2 & x_0-x_3
\end{pmatrix}
\end{align}
so that instead of the usual components $x_\mu = \{x_0,-\vec{x}\}$ each four-vector is described
by its light-cone coordinates $x_\pm = x_0\pm x_3$ and two complex coordinates
in the transverse plane $x_1\pm ix_2$.

The Dirac (quark) spinor $q$ is written as
\begin{align}
q=\begin{pmatrix}\psi_\alpha\\ \bar\chi^{\dot\beta}\end{pmatrix},
&& \bar q=(\chi^\beta,\bar\psi_{\dot\alpha})\,,
\end{align}
where $\psi_{\alpha}$, $\bar \chi^{\dot\beta}$ are two-component Weyl spinors,
$\bar\psi_{\dot\alpha}=(\psi_{\alpha})^\dagger$,
$\chi^{\alpha}=(\bar\chi^{\dot\alpha})^\dagger$.

{}Finally, the gluon strength tensor $F_{\mu\nu}$ and its dual
$\widetilde{F}_{\mu\nu}$ can be decomposed as
\begin{align}
F_{\alpha\beta,\dot\alpha\dot\beta}&=\sigma^\mu_{\alpha\dot\alpha}\sigma^\nu_{\beta\dot\beta}
F_{\mu\nu}=
2\left(\epsilon_{\dot\alpha\dot\beta} f_{\alpha\beta}-
\epsilon_{\alpha\beta} \bar f_{\dot\alpha\dot\beta}
\right)\,,
\notag\\
i {\widetilde F}_{\alpha\beta,\dot\alpha\dot\beta}&=2(\epsilon_{\dot\alpha\dot\beta}
f_{\alpha\beta}+
\epsilon_{\alpha\beta}\bar f_{\dot\alpha\dot\beta})\,.
\end{align}
Here $f_{\alpha\beta}$ and $\bar f_{\dot\alpha\dot\beta}$ are chiral and antichiral
symmetric tensors, $f^*=\bar f$, which belong to $(1,0)$ and $(0,1)$ representations
of the Lorenz group, respectively.  The antisymmetric tensors,
$\epsilon_{\alpha\beta}=\epsilon^{\alpha\beta}$
and $\epsilon_{\dot\alpha\dot\beta}=\epsilon^{\dot\alpha\dot\beta}$
are normalized by $\epsilon_{12}=-\epsilon_{\dot
1\dot 2}=1$ and used for rising and lowering indices
\begin{align}\label{raise}
u^\alpha=\epsilon^{\alpha\beta}u_\beta\,,&& u_\alpha=u^\beta\epsilon_{\beta\alpha}\,,
\notag\\
\bar u^{\dot\alpha}=\bar u_{\dot\beta}\epsilon^{\dot\beta\dot\alpha}\,,&&
\bar u_{\dot\alpha}=\epsilon_{\dot\alpha\dot\beta}\bar u^{\dot\beta}\,.
\end{align}
The contractions $(ab)$ and $(\bar a\bar b)$ are
defined as
\begin{align}
(ab)=a^\alpha b_\alpha = - a_\alpha b^\alpha\,, &&
(\bar a\bar b) = \bar a_{\dot\alpha} \bar b^{\dot\alpha} = - \bar a^{\dot\alpha} \bar b_{\dot\alpha}.
\end{align}
The scalar product of two four-vectors $a$ and $b$ takes the form
$a_\mu b^\mu=\frac12 a_{\alpha\dot\alpha} \bar b^{\dot\alpha\alpha}$.

For convenience, we present the expressions for Dirac matrices
in the spinor basis:
\begin{align}
\gamma^{\mu}=\begin{pmatrix}0&[\sigma^\mu]_{\alpha\dot\beta}\\
                            [\bar\sigma^{\mu}]^{\dot\alpha\beta}&0 \end{pmatrix},&&
\gamma_5=\begin{pmatrix}-\delta_{\alpha}^\beta&0\\
                           0&\delta^{\dot\alpha}_{\dot\beta}  \end{pmatrix}\,,
\end{align}
where 
$\gamma_5=i\gamma^0\gamma^1\gamma^2\gamma^3$.
More expressions and various useful identities can be found in
\cite{Braun:2008ia,Braun:2009vc}.

A covariant generalization of the decomposition in terms of light-cone coordinates
and a transverse plane (\ref{xd}) can be found by observing that any
light-like vector can be represented as a product of two spinors.
In particular one can parameterize the light-like vectors $n$ and $\tilde n$ as follows
\begin{align}
  n_{\alpha\dot\alpha}=\lambda_{\alpha}\bar\lambda_{\dot\alpha}\,,
&&  \tilde n_{\alpha\dot\alpha}=\mu_\alpha\bar\mu_{\dot\alpha}\,, 
\label{2:nlambda}
\end{align}
where $\bar\lambda=\lambda^\dagger$,  $\bar\mu=\mu^\dagger$.
The standard convention $(n\tilde n)=1$ corresponds to the normalization of
auxiliary spinors
\begin{align}
 (\mu\lambda) = (\bar \lambda \bar \mu) = \sqrt{2}\,.
\end{align}
The basis vectors in the
plane transverse to $n,\tilde n$ can be chosen as $\mu_{\alpha}\bar\lambda_{\dot\alpha}$
and $\lambda_{\alpha}\bar\mu_{\dot\alpha}$ and  an arbitrary four-vector
$x$ represented as
\begin{align}
x_{\alpha\dot\alpha}=
z\,\lambda_{\alpha}\bar\lambda_{\dot\alpha}+\tilde z\,\mu_{\alpha} \bar\mu_{\dot\alpha}
+w\,\lambda_{\alpha}\bar\mu_{\dot\alpha}+\bar w\,\mu_{\alpha}\bar\lambda_{\dot\alpha}\,,
\end{align}
where $z$ and $\tilde z$ are real and $w$, $\bar w =w^*$ complex coordinates in the
two light-like directions and the transverse plane, respectively, cf. Eq.~(\ref{xd}).
In particular the spin vectors $s_\mu$ and $\tilde s_\mu$ take the form
\begin{align}\label{}
s_{\alpha\dot\alpha}&=-\frac12
\Bigl\{\lambda_{\alpha}\bar\mu_{\dot\alpha}
s_{\mu\bar\lambda}+
\mu_{\alpha}\bar\lambda_{\dot\alpha} s_{\lambda\bar\mu}\Bigr\}\,,
\notag\\
\tilde s_{\alpha\dot\alpha}&=-\frac{i}2\Bigl\{\lambda_{\alpha}\bar\mu_{\dot\alpha}
s_{\mu\bar\lambda}-
\mu_{\alpha}\bar\lambda_{\dot\alpha} s_{\lambda\bar\mu} \Bigr\}\,,
\end{align}
where $s_{\mu\bar\lambda}=\mu^\alpha
s_{\alpha\dot\alpha}\bar\lambda^{\dot\alpha}\overset{\mathrm{def}}{=}(\mu s\bar\lambda) $.

The ``$+$'' fields  (''good components'')
are defined as the projections onto $\lambda$:
\begin{align*}
\psi_+=\lambda^\alpha\psi_\alpha\,,&&\chi_+=\lambda^\alpha\chi_\alpha\,, &&
f_{++}=\lambda^\alpha\lambda^\beta f_{\alpha\beta}\,,
\nonumber\\
\bar\psi_+=\bar\lambda^{\dot\alpha}\bar\psi_{\dot\alpha}\,,&&
\bar \chi_+=\bar\lambda^{\dot\alpha}\chi_{\dot\alpha}\,, &&
\bar f_{++}=\bar\lambda^{\dot\alpha}\bar\lambda^{\dot\beta} \bar f_{\dot\alpha\dot\beta}\,.
\end{align*}

The scale dependence of the correlation functions (\ref{Qiu1}), (\ref{TDT}) is determined by the
renormalization properties of the corresponding light-ray operators (\ref{TdT}), (\ref{FFF}).
In order to make a connection to the existing results, it is convenient to go over,
for quark-antiquark-gluon operators, to another basis~\cite{Balitsky:1987bk}
\begin{align}\label{Spm}
S^\pm_\rho({z})=g
\bar q(z_1)\big[F_{\rho+}(z_2)\pm i\gamma_5\widetilde{F}_{\rho+}(z_2)\big]\gamma_+ q(z_3)\,.
\end{align}
Here and below we use a shorthand notation $q(z_3) \equiv q(z_3n)$ etc.,
and also $S_{\rho}^{\pm}({z})\equiv S_{\rho}^{\pm}({z_1,z_2,z_3})$.
The operators $S^\pm_\rho$ contribute to the structure function $g_2(x,Q^2)$ for polarized
deep-inelastic scattering and their renormalization properties are studied in
much detail
\cite{Balitsky:1987bk,Ali:1991em,Mueller:1997yk,Koike:1998ry,Derkachov:1999ze,Braun:2000av,Braun:2001qx}.
It is easy to check
that
\begin{align}\label{SvQ}
\tilde s^\rho T_\rho({z})&=\frac12\tilde s^\rho \left(S^+_\rho({z})+S^-_\rho({z})\right)\,,
\notag\\
s^\rho \Delta T_\rho({z})&=-\frac{1}{2}
\tilde s^\rho \left(S^+_\rho({z})-S^-_\rho({z})\right)\,.
\end{align}
The second identity is a consequence of the known representation for the product of
two antisymmetric $\epsilon$--tensors
as the determinant of the $4\times 4$ matrix in terms of the metric tensors $g^{\mu\nu}$.

In the spinor basis  one finds for the operators
$S^\pm_{\alpha\dot\alpha}=S^\pm_\rho\sigma^\rho_{\alpha\dot\alpha}$
the following expressions:
\begin{eqnarray}
S_{\alpha\dot\alpha}^{+}({z})&=&2g\Big[\bar\lambda_{\dot\alpha}\bar\psi_+(z_1)
f_{+\alpha}(z_2)\psi_+(z_3)
\notag\\ & &
+\lambda_{\alpha}\chi_+(z_1)\bar f_{+\dot\alpha}(z_2)\bar\chi_+(z_3)
\Big],
\notag\\
S_{\alpha\dot\alpha}^{-}({z})&=&2g\Big[\bar\lambda_{\dot\alpha}\chi_+(z_1)
f_{+\alpha}(z_2)\bar\chi_+(z_3)
\notag\\ & &
+\lambda_{\alpha}\bar\psi_+(z_1)\bar f_{+\dot\alpha}(z_2)\psi_+(z_3)
\Big].
\end{eqnarray}

Taking into account transformation properties of the quark and gluon fields under charge
conjugation,
\begin{eqnarray}
C\psi C^{-1}=\chi\,, &\qquad& C \bar\psi C^{-1}=\bar \chi\,,
\notag\\
 CfC^{-1}=-f^T,  &\qquad& C\bar fC^{-1}=-\bar f^T,
\end{eqnarray}
where $f=f^{a} t^{a}$, one derives
\begin{equation}
C S^+_\rho(z_1,z_2,z_3) C^{-1}=S^-_\rho(z_3,z_2,z_1).
\end{equation}
Since operators of different $C$-parity do not mix under renormalization, it is
convenient to introduce the $C$-even and $C$-odd combinations
\begin{align}
\mathfrak{S}^\pm({z})=S^+(z_1,z_2,z_3)\pm S^-(z_3,z_2,z_1)\,,
\end{align}
where $S^\pm=\tilde s^\rho S^\pm_\rho$.
Note that these are {\it not} the same combinations as the ones appearing
in Eq.~(\ref{SvQ}) since the quark and antiquark field coordinates are interchanged.
The expressions for the operators $\mathfrak{S}^\pm$ in terms of chiral fields read
\begin{align}\label{Sdef}
 \mathfrak{S}^{\pm}({z})=-\frac{ig}{\sqrt{2}}\left\{
  s_{\mu\bar\lambda}\mathcal{Q}^{\pm}({z})
- s_{\lambda\bar\mu} \widetilde{\mathcal{Q}}^{\pm}({z})
\right\},
\end{align}
where
\begin{align*}\label{}
\mathcal{Q}^{\pm}({z})&=
\bar\psi_+(z_1)f_{++}(z_2)\psi_+(z_3)
\pm \chi_+(z_3) f_{++}(z_2)\bar\chi_+(z_1)\,,
\notag\\
 \widetilde{\mathcal{Q}}^{\pm}({z})&=
\chi_+(z_1)\bar f_{++}(z_2)\bar\chi_+(z_3)
\pm \bar\psi_+(z_3) \bar f_{++}(z_2)\psi_+(z_1)\,.
\end{align*}
It is easy to see that
$\widetilde{\mathcal{Q}}^{\pm}({z})=\pm[\mathcal{Q}^{\pm}({z})]^\dagger$, so
that the $C$-even (``plus'') and $C$-odd (``minus'')
$\mathfrak{S}$-operators are hermitian and antihermitian,
respectively, i.e.
$\left(\mathfrak{S}^{\pm}({z})\right)^\dagger= \pm\mathfrak{S}^{\pm}({z})$.

The operators $\mathfrak{S}^\pm$ have autonomous evolution for the flavor-nonsinglet
sector, whereas in the singlet sector they mix with three-gluon operators $\mathcal{F}^\pm$
of the same $C$-parity:
\begin{eqnarray}\label{Fdef}
\mathcal{F}^\pm({z})&=&-\frac{ig}{\sqrt{2}}C^{abc}_\pm
\Big\{s_{\mu\bar\lambda}\bar f_{++}^{a}(z_1) f_{++}^b(z_2)f_{++}^c(z_3)
\notag\\
&&{}\hspace*{0.8cm}- s_{\lambda\bar\mu} f_{++}^{a}(z_1) \bar f_{++}^b(z_2) \bar f_{++}^c(z_3)\Big\}.
\end{eqnarray}
The color factors $C^{abc}_\pm$ are given in Eq.~(\ref{Cpm}).
In the vector notation this definition corresponds to
\begin{align}\label{}
\mathcal{F}^\pm({z})&=2gC^{abc}_\pm \tilde s^\rho\, (1\mp P_{23} \pm  P_{12})
\notag\\
&\phantom{=2g}
\times F_{+}^{\phantom{+}\nu,a}(z_1)F_{+\rho}^b(z_2) F_{+\nu}^c(z_3)\,,
\end{align}
where $P_{23}$ and $P_{12}$ are the permutation operators acting on the
field coordinates, e.g.
\begin{equation}
 P_{23} F_{+}^{\phantom{\,\,}\nu,a}(z_1)F_{+\rho}^b(z_2) F_{+\nu}^c(z_3)
=
F_{+}^{\phantom{\,\,}\nu,a}(z_1)F_{+\rho}^b(z_3) F_{+\nu}^c(z_2)
\end{equation}
etc.

Taking the nucleon matrix elements of the operators $\mathfrak{S}^\pm$ and $\mathcal{F}^\pm$ one
obtains the corresponding parton correlation functions in momentum space
\begin{align}\label{SFM}
\vev{P, s_T|\mathfrak{S}^\pm({z})|P, s_T}&=2P_+^2\int \mathcal{D}x e^{-iP_+\sum_k x_k z_k}
\,\mathfrak{S}^\pm({x})\,,
\notag\\
\vev{P, s_T|\mathcal{F}^\pm({z})|P,s_T}&=2P_+^3\int \mathcal{D}x e^{-iP_+\sum_k x_k z_k}
\,\mathcal{F}^\pm({x})\,.
\end{align}
Here $x =\{x_1,x_2,x_3\}$, the integration measure $\mathcal{D}x$ is defined in Eq.~(\ref{measure}).
We use the same notation for the correlation functions as for the corresponding
operators which, hopefully, will not result in a confusion.
Note that although definitions of the
operators $\mathfrak{S}^\pm$ and $\mathcal{F}^\pm$,
Eqs.~(\ref{Sdef}) and (\ref{Fdef}), involve the spin-vector $s^\rho$,
the dependence on $s^\rho$ actually drops out in the matrix elements~(\ref{SFM}).

Hermiticity and $C$-invariance imply that (cf.~\cite{Qiu:1998ia})
\begin{align}\label{C-inv}
(\mathfrak{S}^\pm({x}))^*&=\pm\mathfrak{S}^\pm(-{x})\,, &
\mathfrak{S}^\pm({x})&=\pm \mathfrak{S}^\pm(-{x})\,,
\notag\\
(\mathcal{F}^\pm({x}))^*&=\mp\mathcal{F}^\pm(-{x})\,, &
\mathcal{F}^\pm({x})&=\mp \mathcal{F}^\pm(-{x})\,.
\end{align}
This means, in particular, that the correlation functions
$\mathfrak{S}^\pm({x})$ and $\mathcal{F}^\pm({x})$
are real functions. Notice also that the function $\mathcal{F}^-$ is symmetric and
$\mathcal{F}^+$ antisymmetric under the interchange of the last two arguments:
\begin{align*}
\mathcal{F}^\pm(x_1,x_2,x_3)=\mp \mathcal{F}^\pm(x_1,x_3,x_2)\,.
\end{align*}
The six correlation functions defined in Eq.~(\ref{TDT})
(or, equivalently, Eq.~(\ref{Qiu1})) can be written in terms of two independent
quark-antiquark-gluon functions $\mathfrak{S}^\pm({x})$ and
two three-gluon functions $\mathcal{F}^\pm({x})$ as follows:
\begin{align}\label{relation11}
T_{\bar q F q}({x})&=\frac14\Big[(1+P_{13})\mathfrak{S}^+({x})+
(1-P_{13})\mathfrak{S}^-({x})\Big],
\notag\\
\Delta T_{\bar q F q}({x})&=-\frac14\Big[(1-P_{13})\mathfrak{S}^+({x})+
(1+P_{13})\mathfrak{S}^-({x})\Big],
\notag\\
T_{3F }^\pm({x})&=\frac12(1\mp P_{13})\mathcal{F}^{\pm}({x})\,,
\notag\\
\Delta T_{3F }^\pm({x})&=-\frac12(1\pm P_{13})\mathcal{F}^{\pm}({x})\,.
\end{align}
Here $T_{\bar q F q}(x)\equiv T_{\bar q F q}(x_1,x_2,x_3)$, etc., and $P_{ik}$
are the permutation operators for the corresponding momentum fractions,
e.g. $P_{12}\mathcal{F}^{\pm}(x_1,x_2,x_3) \equiv \mathcal{F}^{\pm}(x_2,x_1,x_3)$.
As follows from (\ref{relation11}) the functions $T_{3F }^\pm({x})$ and
$\Delta T_{3F}^\pm({x})$ are not independent,
\begin{align}\label{TDeltaT}
\Delta T_{3F}^\pm(x_1,x_2,x_3)=\pm\Big[T_{3F}^\pm(x_1,x_3,x_2)- T_{3F}^\pm(x_2,x_1,x_3)\Big]
\end{align}
It follows from~(\ref{C-inv}) and (\ref{relation11}) that the correlation functions
satisfy the following symmetry relation
\begin{align}\label{symmetry5}
T_{\bar q F q}(x_1,x_2,x_3)&=T_{\bar q F q}(-x_3,-x_2,-x_1)\,,
\notag\\
\Delta T_{\bar q F q}(x_1,x_2,x_3)&=-\Delta T_{\bar q F q}(-x_3,-x_2,-x_1)\,,
\notag\\
T_{3 F }^\pm(x_1,x_2,x_3)&=T_{3 F}^\pm(-x_3,-x_2,-x_1)\,,
\notag\\
\Delta T_{3F }^\pm(x_1,x_2,x_3)&=-\Delta T_{3 F }^\pm(-x_3,-x_2,-x_1)\,.
\end{align}
Authors of~\cite{Kang:2008ey} also introduce symmetrized quark-antiquark-gluon parton distributions
\begin{align}\label{}
\mathcal{T}_{q,F}(x,x')&=
\frac12\left(\widetilde{\mathcal{T}}_{q,F}(x,x')+\widetilde{\mathcal{T}}_{q,F}(x',x) \right),
\notag\\
\mathcal{T}_{\Delta q,F}(x,x')&=
\frac12\left(\widetilde{\mathcal{T}}_{\Delta q,F}(x,x')-
\widetilde{\mathcal{T}}_{\Delta q,F}(x',x) \right).
\end{align}
As follows from Eq.~(\ref{symmetry5}) such a symmetrization is not necessary since
$\widetilde{\mathcal{T}}_{q,F}(x,x') = \widetilde{\mathcal{T}}_{q,F}(x',x)$
and $\Delta\widetilde{\mathcal{T}}_{q,F}(x,x') = -\Delta\widetilde{\mathcal{T}}_{q,F}(x',x)$.
In the expressions given below we drop the ``tilde'' notation for these functions.

\section{Flavor-nonsinglet evolution}
The flavor-nonsinglet light-ray operators $\mathfrak{S}^+$ and
$\mathfrak{S}^-$ satisfy the same evolution equation,
\begin{align}\label{RG1}
\left(\mu\frac{\partial}{\partial\mu}+\beta(g)\frac{\partial}{\partial g}+\frac{\alpha_s}{2\pi}
\mathbb{H}\right)\mathfrak{S}^\pm=0\,,
\end{align}
where $\mathbb{H}$ is an integral operator. The explicit expression for $\mathbb{H}$
(at one-loop) can be restored from the corresponding result for $S^\pm$ in \cite{Balitsky:1987bk}
(see also \cite{Mueller:1997yk,Braun:2000av,Braun:2001qx}).
Expanding this equation at short distances $z_i\to0$ one reproduces the mixing matrix
for the twist-three local operators~\cite{Bukhvostov:1984as,Ji:1990br,Kodaira:1996md,Koike:1998ry}.

Alternatively, the answer for $\mathbb{H}$ can be obtained by simple algebra
from the known expressions for the one-loop two-particle
kernels~\cite{Bukhvostov:1985rn,Braun:2009vc}. This technique is general and applicable to
arbitrary twist-three (and twist-four \cite{Braun:2009vc}) evolution equations. In the next
Section we will use this approach for a more complicated case of flavor-singlet operators
so it makes sense to explain the details on the present (simpler) example.

The starting observation is that to one-loop accuracy any contribution to
the evolution can only involve two partons; hence
$\mathbb{H}$ can be represented as a sum of two-particle kernels.
 Schematically
\begin{equation}
\mathbb{H}\,\bar\psi_+ \!f_{++}\psi_+ =
t^a_{ik}\! \Big\{\wick{2}{<1{\overline{\psi}}^i_+>1 f^a_{++}\psi^k_+}
+\wick{2}{{\overline{\psi}}^i_+ <1 f^a_{++} >1 \psi^k_+}
+\wick{2}{<1 {\overline{\psi}}^i_+ f^a_{++} >1 \psi^k_+}\! \Big\}
\end{equation}
where the contractions correspond to the sum of relevant Feynman diagrams (in light-cone
gauge). The corresponding expressions were derived originally by
Bukhvostov, Frolov, Lipatov and Kuraev (BFLK)~\cite{Bukhvostov:1985rn}.
The complete list of the BFLK kernels for arbitrary chiral fields is given
in Ref.~\cite{Braun:2009vc} so that it only remains
to contract the color indices. For the reader's convenience we collect all
the kernels in Table~\ref{table1} in the Appendix.

After a simple algebra one gets
\begin{align}\label{HNS}
\mathbb{H}=N_c\, \mathbb{H}_0-\frac{1}{N_c}\mathbb{H}_1-3 C_F\,,
\end{align}
where, in notation of Ref.~\cite{Braun:2009vc},
\begin{align}\label{Hpw}
 \mathbb{H}_0&=\widehat{\mathcal{H}}_{12} +\widehat{\mathcal{H}}_{23}-
    2\mathcal{H}_{12}^+\,,\notag\\
\mathbb{H}_1&=\widehat{\mathcal{H}}_{13}-\mathcal{H}_{13}^+-P_{23}\mathcal{H}_{23}^{e,(1)}+
2\mathcal{H}_{12}^-\,.
\end{align}
Here $\mathcal{H}_{ik}$ are two-particle integral operators that act on the
light-cone coordinates of the $i$-th and $k$-th partons:
\begin{equation}
 [\mathcal{H}_{ik}\phi](z_i,z_k)=\int dz'_i dz'_k\,
\mathcal{H}(z_i,z_k|z'_i,z'_k)\,\phi(z'_i,z'_k)\,.
\end{equation}
These kernels are $SL(2)$-invariant and depend on the conformal spins of partons
that they are acting on. The corresponding values are $j=1$ for quarks and $j=3/2$
for gluons (for the ``plus'' components) so one has to use $j_1=1$, $j_2=3/2$, $j_3=1$.
{}One obtains, for example,
\begin{align*}
[\mathcal{H}_{13}^+\mathfrak{S}^+](z_1,z_2,z_3)=\int_0^1d\alpha\int_0^{\bar\alpha}d\beta\,
\mathfrak{S}^+(z_{13}^\alpha,z_2,z_{31}^\beta)\,,
\end{align*}
where
$$z_{ik}^\alpha=z_i\bar\alpha+z_k\alpha\,, \qquad \bar\alpha = 1-\alpha\,.$$

Taking the nucleon matrix element of Eq.~(\ref{RG1}) one obtains the evolution equation
for the corresponding parton distribution function. Technically,
this corresponds to going over from coordinate to the momentum fraction space
$\{z_1,z_2,z_3\}\to \{x_1,x_2,x_3\}$.
Thanks to energy conservation,
two-particle kernels in momentum fraction space can be written in the following
generic form
\begin{align}\label{MV}
[\mathcal{H}_{ik}\varphi](x_i,x_k)&=\int_{-\infty}^\infty \mathcal{D}x'\,
\mathcal{H}(x_i,x_k|x'_i,x'_k)\,\varphi(x'_i,x'_k)\,,
\end{align}
where $\mathcal{D}x'=dx'_idx'_k \delta(x_i+x_k-x'_i-x'_k)$. It is assumed that
restrictions on integration regions over $x'_1,x'_2$ come from support properties of the kernels
and the parton distributions; the corresponding
expressions are collected in the Appendix.

The last step, using the first two relations in Eq.~(\ref{relation11}) we obtain
\begin{widetext}
\begin{eqnarray}
\mu \frac{d}{d\mu} T_{\bar q F q}(x) &=&
- \frac{\alpha_s}{4\pi}\Big(\mathbb{H}+P_{13}\,\mathbb{H}\,P_{13}\Big)  T_{\bar q F q}(x)
+  \frac{\alpha_s}{4\pi}\Big(\mathbb{H}-P_{13}\,\mathbb{H}\,P_{13}\Big) \Delta T_{\bar q F q}(x)\,,
\nonumber\\
\mu \frac{d}{d\mu} \Delta T_{\bar q F q}(x) &=&
- \frac{\alpha_s}{4\pi}\Big(\mathbb{H}+P_{13}\,\mathbb{H}\,P_{13}\Big) \Delta T_{\bar q F q}(x)
+  \frac{\alpha_s}{4\pi}\Big(\mathbb{H}-P_{13}\,\mathbb{H}\,P_{13}\Big) T_{\bar q F q}(x)\,,
\end{eqnarray}
\end{widetext}
which is our final result. The ``Hamiltonian'' $\mathbb{H}$ is defined in Eqs.~(\ref{HNS}), (\ref{Hpw}).
Explicit expressions for the two-particle kernels
$\widehat{\mathcal{H}}_{ik}$, $\mathcal{H}_{ik}^+$, $\mathcal{H}_{ik}^{e,(1)}$ and $\mathcal{H}_{ik}^-$
are given in the Appendix. Note that when applying two-particle kernels to a three-particle parton
distribution one has to treat the latter as a function of three independent variables;
the condition $x_1 + x_2 + x_3 =0$ is applied afterwards.

\begin{figure}[t]
\includegraphics[width=6cm]{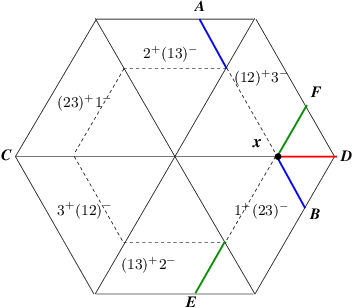}
\caption{Integration regions in Eq.~(\ref{KQe}).
\label{figure3}}
\end{figure}

To compare our result with the calculation in Ref.~\cite{Kang:2008ey} we have to change
notation to
\begin{align}
{\mathcal{T}}_{q,F}(x,x')&\equiv
T_{\bar q F q}(-x',x'-x,x)\,,
\notag\\
{\mathcal{T}}_{\Delta q,F}(x,x')&\equiv  \Delta T_{\bar q F q}(-x',x'-x,x)\,,
\end{align}
which are (anti)symmetric functions under permutation $x\leftrightarrow x'$,
${\mathcal{T}}_{q,F}(x,x')={\mathcal{T}}_{q,F}(x',x)$ and
$\Delta{\mathcal{T}}_{q,F}(x,x')=-\Delta{\mathcal{T}}_{q,F}(x',x)$.
Using explicit expressions for the two-particle kernels and taking the gluon-pole limit $x'=x$ ($x_2=0$)
we obtain (it is assumed that $x>0$)
\begin{widetext}
\begin{align}\label{KQe}
\mu\frac{d}{d\mu}
\mathcal{T}_{q,F}(x,x)&=
\frac{\alpha_s}{\pi}\Biggl\{\int_x^1 \frac{d\xi}{\xi}\Biggl[P_{qq}(z) \mathcal{T}_{q,F}(\xi,\xi)
+\frac{N_c}2\left(\frac{(1+z)\mathcal{T}_{q,F}(x,\xi)-(1+z^2)\mathcal{T}_{q,F}(\xi,\xi)}{1-z}
-\mathcal{T}_{\Delta q,F}(x,\xi)\right)\Biggr]\notag\\
&-N_c \mathcal{T}_{q,F}(x,x)
+\frac1{2N_c}\int_x^{1}\frac{d\xi}{\xi}
\biggl[(1-2z)\mathcal{T}_{q,F}(x,x-\xi)-\mathcal{T}_{\Delta q,F}(x,x-\xi)\biggr]\Biggr\}\,,
\end{align}
\end{widetext}
where $z=x/\xi$,
\begin{align}\label{Pqq}
P_{qq}(z)=C_F\left[\frac{1+z^2}{(1-z)_+}+\frac32\delta(1-z)\right]
\end{align}
and
\begin{align*}\label{}
\int_x^1dz\frac{f(z)}{(1-z)_+}=\int_x^1dz\frac{f(z)-f(1)}{1-z}+f(1)\log(1-x)\,.
\end{align*}

The partonic interpretation of different contributions in Eq.~(\ref{KQe}) is illustrated
in Fig.~\ref{figure3}. Note that the condition of zero gluon momentum $x_2=0$ corresponds to the
choice of $\vec{x}$ on the (positive) horizontal axis.
The first term $\sim \mathcal{T}_{q,F}(\xi,\xi)$ on the r.h.s. of Eq.~(\ref{KQe})
corresponds to the integration over the horizontal line segment {\small\bf xD} shown in red in Fig.~\ref{figure3}.
The terms in  $\mathcal{T}_{q,F}(x,\xi)$ are due to the integration
along the blue, {\small\bf AB}, and green, {\small\bf EF}, lines in the regions
$1^+(23)^-$ and $(12)^+3^-$,
and  the ones in $\mathcal{T}_{q,F}(x,x-\xi)$ correspond to the contributions
along the same lines
in the regions $2^+(13)^-$ and $(13)^+ 2^-$, respectively.
By construction, the function  $\mathcal{T}_{q,F}(x,x')$ is symmetric and $\mathcal{T}_{\Delta q,F}(x,x')$
antisymmetric under the interchange of arguments,  $x\leftrightarrow x'$.
In Fig.~\ref{figure3} this corresponds to a reflection symmetry around the horizontal axis.

\begin{figure}[t]
\includegraphics[width=3cm]{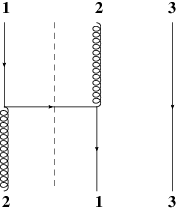}
\caption{The exchange diagram}
\label{figure4}
\end{figure}

The evolution equation (\ref{KQe}) differs from the corresponding result by Kang and Qiu
(see Eq.~(99) in  \cite{Kang:2008ey}) by the two extra terms in the second line.
The last term, proportional to $1/N_c$, originates from
the kernels  $P_{23}\mathcal{H}_{23}^{e,(1)}$ and $\mathcal{H}_{12}^- $ in Eq.~(\ref{Hpw}).
On the diagrammatic level this contribution corresponds to the ``exchange'' diagrams of the type
shown in Fig.~\ref{figure4} which correspond to mixing of the regions that have different
partonic interpretation: $2^+(13)^- \leftrightarrow 1^+(23)^-$,
or $(13)^+ 2^- \leftrightarrow (12)^+ 3^-$. Similar contributions have been discussed in a somewhat
different context in Ref.~\cite{Koike:2009yb}.

The other difference is the extra term $-N_c \mathcal{T}_{q,F}(x,x)$ in the second line in Eq.~(\ref{KQe})
for which we do not see any obvious explanation.
 The origin of this term in our calculation can be traced to the kernels
$\widehat{\mathcal{H}}_{12}$ and $\widehat{\mathcal{H}}_{23}$.
E.g. the second term in the expression for $\widehat{\mathcal{H}}_{12}$,
Eq.~(\ref{A:Hhat}), gives
\begin{eqnarray}
\lefteqn{
\mu\frac{d}{d\mu}\mathcal{T}_{q,F}(x,x+x_2)\,\stackrel{x_2\to 0+}{=}}
\nonumber\\&=&\ldots
-\frac{\alpha_{s} N_c}{4\pi} \int_{x_2}^\infty \frac{x_2 dx'_2}{x'_2(x'_2-x_2)}
\nonumber\\
&&\times\Big[\mathcal{T}_{q,F}(x,x+x_2)
-\frac{x_2}{x'_2}\mathcal{T}_{q,F}(x,x+x'_2)\Big].
\label{subtle}
\end{eqnarray}
Because of an overall $x_2$ factor, it is tempting to put this contribution
to zero in the $x_2\to0$ limit. However, it is easy to see that in the same limit
the integral becomes linearly divergent so that at the end a finite contribution
arises. This term can easily be missed if the gluon momentum is put to zero at the
beginning of the calculation.

It is instructive to analyze Eq.~(\ref{KQe}) in the large--$N_c$ limit. Neglecting $1/N_c$
terms in~(\ref{KQe}) one is left with
\begin{align}\label{largeNc}
&\mu\frac{d}{d\mu}
\mathcal{T}_{q,F}(x,x)=
\frac{\alpha_s N_c}{2\pi}\Biggl\{-\mathcal{T}_{q,F}(x,x)
\notag\\
&+\int_x^1 \frac{d\xi}{\xi}
\Biggl[\left(\bar P_{qq}(z)+{z}\right) \mathcal{T}_{q,F}(x,\xi)-
\mathcal{T}_{\Delta q,F}(x,\xi)\Biggr]\Biggl\}\,,
\end{align}
where $\bar P_{qq}(z)$ is obtained from $P_{qq}(z)$, Eq.~(\ref{Pqq}), omitting the $C_F$ factor.
One sees that the scale dependence of $\mathcal{T}_{q,F}(x,x)$
is determined by $\mathcal{T}_{q,F}(x,\xi)$ in the region $x\leq \xi\leq 1$,
which corresponds to the contribution of the blue and green line segments
{\small\bf xB} and {\small\bf xF} in Fig.~\ref{figure3}.
It receives no contribution from the ``diagonal'' region $ \mathcal{T}_{q,F}(\xi,\xi)$
(the red line segment {\small\bf xD} in Fig.~\ref{figure3}),
the corresponding contributions cancel out between
the first and the second terms in ~Eq.~(\ref{KQe}) (to the $\mathcal{O}(1/N_c)$ accuracy).
By this reason the conclusion in Ref.~\cite{Kang:2008ey} that the evolution of the function
$\mathcal{T}_{q,F}(x,x)$ mainly follows  a pattern determined by the quark splitting
function can be misleading.

The evolution does simplify, however, in the large $x$ limit in which case the integration
regions shrink to a point. One obtains
\begin{equation}
\mu\frac{d}{d\mu}\mathcal{T}_{q,F}(x,x) =
\frac{\alpha_s}{\pi}\int_x^1\frac{d\xi}{\xi}P^{NS,z\to1}_{q,F}(z)\mathcal{T}_{q,F}(\xi,\xi)\,,
\end{equation}
where, retaining singular terms at $z\to 1$ only
\begin{equation}
P^{NS,z\to1}_{q,F}(z) =  2C_F\left[\frac{1}{(1-z)_+} + \frac34\delta(1-z)\right] - N_c\delta(1-z)\,.
\end{equation}
This result can be compared with the evolution of the usual $F_1(x,Q^2)$ structure
function which involves, to the same accuracy
\begin{equation}
P^{NS,z\to1}_{qq}(z) = 2C_F\left[\frac{1}{(1-z)_+} + \frac34\delta(1-z)\right],
\end{equation}
and the twist-three contribution to the structure function $g_2(x,Q^2)$ \cite{Ali:1991em,Braun:2000av}
\begin{equation}
P^{NS,z\to1}_{g_2}(z) = 2C_F\left[\frac{1}{(1-z)_+} + \frac34\delta(1-z)\right] -\frac{N_c}{2}\delta(1-z)\,.
\end{equation}
The last term in $\delta(1-z)$ is written in the large-$N_c$ limit.
The contributions $\sim 1/(1-z)_+$ are the same in all three cases, which indicates that all three
functions $\mathcal{T}_{q,F}(x,x;Q^2)$, $F_1(x,Q^2)$ and $g^{tw.-3}_2(x,Q^2)$ may have the same
functional dependence on the Bjorken variable $x$ in the $x\to1$ limit. Different
terms $\sim\delta(1-z)$ suggest, on the other hand, that twist-three functions are suppressed
at large scales $Q^2$ compared to the twist-two distribution
as
\begin{eqnarray}
 \mathcal{T}_{q,F}(x,x;Q^2)/F_1(x,Q^2) &\sim& \left(\frac{\alpha_s(Q)}{\alpha_s(\mu_0)}\right)^{2N_c/b_0},
\nonumber\\
g^{tw.-3}_2(x,Q^2)/F_1(x,Q^2) &\sim& \left(\frac{\alpha_s(Q)}{\alpha_s(\mu_0)}\right)^{N_c/b_0}.
\end{eqnarray}
The suppression of  $g^{tw.-3}_2(x,Q^2)$ compared to  $F_1(x,Q^2)$ exactly corresponds to the
gap between  the lowest anomalous dimension in the spectrum of twist-three operators
and the usual twist-two anomalous dimension. For the function $\mathcal{T}_{q,F}(x,x;Q^2)$
we predict a stronger suppression which translates to scaling violation in SSA.
This result can be phenomenologically relevant.

\section{Flavor-singlet evolution}
In the flavor-singlet sector one has to take into account mixing
between the quark-antiquark-gluon and three-gluon operators with
the same $C$-parity. Namely, $\mathfrak{S}^+$ gets mixed with $\mathcal{F}^+$ and
$\mathfrak{S}^-$ with~$\mathcal{F}^-$. For each case, the evolution equation takes the matrix form
\begin{align}\label{matrix}
\left(\mu\frac{\partial}{\partial\mu}+\beta(g)\frac{\partial}{\partial g}+
\frac{\alpha_s}{4\pi}\mathbb{H}^\pm\right)\begin{pmatrix}\mathfrak{S}^{\pm}\\
\mathcal{F}^{\pm} \end{pmatrix} =0\,,
\end{align}
where
\begin{align}
\mathbb{H}^\pm=
\begin{pmatrix}\mathbb{H}_{QQ}^{\pm} & \mathbb{H}^{\pm}_{Q F}\\
\mathbb{H}^{\pm}_{F Q} &\mathbb{H}_{FF}^{\pm}
 \end{pmatrix}\,.
\end{align}
In what follows we assume that the quark-antiquark-gluon
flavor-singlet operator is defined including the sum over
$n_f$ light flavors, $\bar q F q = \bar u F u + \bar d F d +\ldots$.

For the diagonal entries we obtain
\begin{align}\label{}
\mathbb{H}_{QQ}^{+}=\mathbb{H}+4 n_f\mathcal{H}^d_{13}\,,
&&
\mathbb{H}_{QQ}^-=\mathbb{H}\,,
\end{align}
where $\mathbb{H}$ is given by Eq.~(\ref{HNS}) and
\begin{align}\label{HFF}
\mathbb{H}_{FF}^{\pm}&=N_c\Big(
\widehat{\mathcal{H}}_{12} + \widehat{\mathcal{H}}_{23}+\widehat{\mathcal{H}}_{31}
-4(\mathcal{H}_{12}^+ + \mathcal{H}_{13}^+)
\notag\\
&-2(\widetilde{\mathcal{H}}_{12}^+ +
\widetilde{\mathcal{H}}_{13}^+)\pm 6(\mathcal{H}_{12}^-+\mathcal{H}_{13}^-)\Big)
-b_0\,,
\end{align}
with $b_0=\dfrac{11}3 N_c-\dfrac23 n_f$. The off-diagonal entries in coordinate space
take the form
\begin{align}\label{Off}
\mathbb{H}_{QF}^{\pm}&=-in_fz_{13}\biggl\{\mathcal{H}_{13}^++\widetilde{\mathcal{H}}_{13}^+
\mp 2\mathcal{H}_{13}^-
\biggr\}\,,
\notag\\
\mathbb{H}_{FQ}^{+}&=iN_c (1-P_{23}) \frac{1}{z_{13}}\Big[2\mathcal{H}_{13}^+P_{13}+1\Big ]\Pi_0\,,
\notag\\
\mathbb{H}_{FQ}^{-}&=-i\frac{N_c^2-4}{N_c}(1+P_{23})\frac{1}{z_{13}}
\Big[2\mathcal{H}_{13}^+P_{13}-1\Big ].
\end{align}
The corresponding expressions in momentum space are given in the
Appendix, Eqs.~(\ref{A:QF})--(\ref{A:last}).

The evolution equations for conventional $T_{\bar q F q}$, $\Delta T_{\bar q F q}$, $T_{3F }^\pm({x})$
$\Delta T_{3F }^\pm$ can readily be obtained from Eq.~(\ref{matrix}) by symmetrization
in the arguments, as specified in Eq.~(\ref{relation11}).

In the limit of zero gluon momentum our result for the evolution of ${T}_{q,F}(x,x) \equiv
T_{\bar q F q}(-x,0,x)$ differs from the corresponding expression in Eq.~(107) of
 \cite{Kang:2008ey} by the same two terms as in the nonsinglet case; the terms in
$P_{qg}$ (in our calculation due to $\mathbb{H}_{QF}$) coincide.

To compare our results for three-gluon distributions we introduce the functions
\begin{align}\label{}
T_F^{\pm}(x,x')&=\frac{1}{x}
{T}_{3F}^\pm(-x',x-x',x)\,,
\notag\\
\Delta T_F^{\pm}(x,x')&=\frac{1}{x}
\Delta{T}_{3F}^\pm(-x',x-x',x)\,,
\end{align}
which coincide with $ T_{G,F}^{(f)}(x,x')$,
$T_{G,F}^{(d)}(x,x')$, $ T_{\Delta G,F}^{(f)}(x,x')$,
$T_{\Delta G,F}^{(d)}(x,x')$ defined in \cite{Kang:2008ey}, respectively.
We remind that the $\Delta T_F^{\pm}(x,x')$ distributions and hence
$ T_{\Delta G,F}^{(f)}(x,x')$, $T_{\Delta G,F}^{(d)}(x,x')$ can be expressed in terms of
$T_F^{\pm}(x,x')$ alias $T_{G,F}^{(d)}(x,x')$, $ T_{ G,F}^{(f)}(x,x')$, so that they
do not need to be considered separately.
After some algebra one obtains the following equations
for $T_F^\pm(x,x)$~\cite{erratum}
\begin{widetext}
\begin{align}\label{TF}
\mu\frac{d}{d\mu}T_F^\pm(x,x)&=
\frac{\alpha_s N_c}{\pi}\biggl(-T_F^\pm(x,x)
+\int_x^1\frac{d\xi}{\xi}\biggl\{2\bar P_{gg}(z)T_F^\pm(\xi,\xi) +\frac{z}{1-z}
\left[T_F^\pm(\xi,x)-T_F^\pm(\xi,\xi)\right]
\notag\\
&
\phantom{=}
-(1-z)\left(z+\frac1z\right)T_F^\pm(\xi,\xi)
+\frac{1+z}{2}\Big[T_F^\pm(x,\xi)-\Delta T_F^\pm(x,\xi)\Big]
\notag\\
&\phantom{=}
\mp\frac12 (1-z)\Big[T_F^\pm(x,x-\xi)
-
\Delta T_F^\pm(x,x-\xi)\Big]
%
+\frac12A^{\pm}\biggl(\bar P_{gq}(z)\Big[{\mathcal{T}}_{q,F}(\xi,\xi)\pm {\mathcal{T}}_{q,F}(-\xi,-\xi)\Big]
\notag\\
&
-
\frac{2-z}{z}\Big[\mathcal{T}_{q,F}(x-\xi,-\xi)\pm \mathcal{T}_{q,F}(\xi,\xi-x)\Big]
+\Big[\Delta \mathcal{T}_{q,F}(x-\xi,-\xi)\mp \Delta \mathcal{T}_{q,F}(\xi,\xi-x)
\Big]
\biggl)\biggr\}\biggr),
\end{align}
\end{widetext}
where
\begin{align}
A^+=1,&& A^-=\frac{N_c^2-4}{N_c^2}
\end{align}
and
\begin{align}\label{}
\bar P_{gg}(z)&=\frac{z}{(1-z)}_++\frac{1-z}{z}+z(1-z)+\frac{b_0}{4N_c}\delta(1-z)\,,
\notag\\
\bar P_{gq}(z)&=\frac{1+(1-z)^2}{z}\,.
\end{align}
Our result does not agree with that of Kang and Qiu \cite{Kang:2008ey}, Eqs.~(109),~(110)
in an overall sign in front of the $T_{\Delta G, F}^{(f(d))}$ distribution
(which may, however, be an artifact of
a different sign convention for $\epsilon_{\mu\nu\alpha\beta}$), and two extra terms
(up to the $\alpha_s N_c/\pi$ factor)
\begin{align}\label{miss2}
T_F^\pm(x,x)\pm\int_x^1\frac{d\xi}{\xi}\frac{1-z}{2}
(T_F^\pm(x,x-\xi)-\Delta T_F^\pm(x,x-\xi))\,
 \end{align}
which seem to have the same origin as the extra contributions that we also have
for the flavor-nonsinglet distributions:
The first term in (\ref{miss2}) originates from contributions of the type
in~Eq.~(\ref{subtle}) that involve a subtlety in taking the $x_2\to 0$ limit, and
the second one corresponds to the contribution of ``exchange'' type diagrams,
the kernel $\mathcal{H}_{12}^-$ in~(\ref{HFF}), that give rise to mixing of regions
with different partonic interpretation.

Closing this section, we want to stress that the scale dependence has to be studied
using complete evolution equations for the three-particle parton distributions,
Eq.~(\ref{matrix}). Using the gluon-pole projected equations (\ref{TF}) with a
certain ansatz for the ``off-diagonal'' correlation functions
$\mathcal{T}^{(f,d)}_{G,F}(x,x')\equiv {T}^{\pm}_{3F}(-x',x'-x,x)$ may be misleading
as they are  modified by the evolution themselves. We note in passing that
the particular ansatz proposed in  \cite{Kang:2008ey}
\begin{align}\label{}
\mathcal{T}^{(f,d)}_{G,F}(x_1,x_2)&=\frac12\biggl[\mathcal{T}^{(f,d)}_{G,F}(x_1,x_1)
\notag\\
&\phantom{=\frac12}+\mathcal{T}^{(f,d)}_{G,F}(x_2,x_2)\biggr]e^{-{(x_1-x_2})^2/2\sigma^2}
\end{align}
and $\mathcal{T}^{(f,d)}_{\Delta G,F}(x_1,x_2)=\Delta{T}^{\pm}_{3F}(-x',x'-x,x)=0$,
is inconsistent with the constraint~(\ref{TDeltaT}).

\section{Conclusions}

We have given a complete reanalysis of the scale dependence of twist-three  three-particle
correlation functions that are relevant for calculations of single transverse spin asymmetries
in the framework of collinear factorization. The calculation is done using the
two-particle kernels for the renormalization of light-ray operators in the spinor basis,
which are available from Ref.~\cite{Braun:2009vc}. Evolution equations are derived
for arbitrary parton momentum fractions, for the flavor-nonsiglet
quark-antiquark-gluon distribution,
Eqs.~(\ref{RG1}), (\ref{HNS}), (\ref{Hpw}), and for
the mixing matrix of the flavor-singlet quark-antiquark-gluon and three-gluon
distributions with both positive and negative $C$-parity, Eqs.~(\ref{matrix})--(\ref{HFF})
and (\ref{A:QF})--(\ref{A:last}). Specializing to the case of zero gluon momentum
we have compared our results with the recent calculation in Ref.~\cite{Kang:2008ey}.
There are two terms where we disagree, and their origin could be identified.
As a byproduct of our calculation we predict logarithmic scaling violation in the SSA
at large values of Bjorken $x$ which may be phenomenologically relevant.
Numerical studies of the evolution effects on realistic models of parton distributions
will be considered elsewhere.

\section*{Acknowledgements}
We are grateful to  Yu.~Koike, K.~Tanaka and W.~Vogelsang for correspondence which
initiated this study and useful comments. This work was supported by the German Research
Foundation (DFG), grant 9209282, RNP grant 2.1.1/1575 and the RFFI grants 07-02-92166,
09-01-93108.


\appendix
\renewcommand{\theequation}{\Alph{section}.\arabic{equation}}

\section*{Appendices}

\section{BFLK kernels in momentum representation}\label{App:A}

Gauge-invariant $N$-particle quasipartonic light-ray operators can be defined as a product of
``plus'' fields
\begin{eqnarray}
\lefteqn{\mathcal{O}(z_1,\ldots,z_N)= C X(z_1)\otimes\ldots\otimes X(z_N)\equiv}
\nonumber\\
&&\hspace*{-3mm} \equiv C_{i_1\ldots i_N} \big([0,z_1]X(z_1)\big)^{i_1}
\ldots \big([0,z_N]X(z_N)\big)^{i_N}\!\!\!,
\label{3:O}
\end{eqnarray}
where $X(z_k)= \{\psi_+,\bar\psi_+,\chi_+,\bar\chi_+,f_{++},\bar f_{++}\}$, $[0,z_k]$ are
Wilson lines in the appropriate representation of the gauge group,
$i_1,\ldots,i_N$ are color indices and $C_{i_1\ldots i_N}$ is an invariant color tensor such that
\begin{align}
\label{3:S-inv}
[(t_1)^a_{k_1i_1} + (t_2)^a_{k_2i_2} +\ldots+ (t_N)^a_{k_Ni_N}]S_{i_1,\ldots,i_N}=0\,.
\end{align}
Here and below it is implied that the generators $t^a$ are taken in the appropriate
representation,
\begin{align}
t^a X=
\begin{cases}\label{3:generators}
(t^a\psi)^i=T^a_{ii'}\psi^{i'}  & \text{for quarks}\quad \psi,\bar\chi
\\
(t^a \bar\psi)^i=-T^a_{i'i}\bar\psi^{i'}&
  \text{for antiquarks}\quad \bar\psi,\chi
\\
(t^a f)^b=i\,\mathrm{f}^{bab'}\! f^{b'}
 &  \text{for gluons }\quad  f,\bar f
\end{cases}
\end{align}	
where $T^a$ are the generators in fundamental representation.
The condition in Eq.~(\ref{3:S-inv})
ensures that $\mathcal{O}(z_1,\ldots,z_N)$ is a color singlet.

{}For each $N$, the set of quasipartonic operators with the
same quantum numbers is closed under renormalization \cite{Bukhvostov:1985rn}.
A renormalized quasipartonic operator
is written as
\begin{align}
[\mathcal{O}_i(X)]_R=\mathbb{Z}_{ik}\mathcal{O}_{k}(X_0)\,,
\end{align}
where $X_0=Z_X X$ is the bare field. Renormalized operators satisfy the RG equation
\begin{align}
\left(\mu\frac{\partial}{\partial\mu}+\beta(g)\frac{\partial}{\partial g}
+{\gamma}_{ik}\right)
[O_k(X)]_R=0\,.
\end{align}
Here $\beta(g)$ is the (QCD) beta function and
$$
\gamma=-\mu\dfrac{d}{d\mu} \mathbb{Z}\, \mathbb{Z}^{-1}
$$
is the matrix of anomalous dimensions.
To the one-loop accuracy one obtains in dimensional regularization ($D=4-2\epsilon$)
\begin{align}
\mathbb{Z}=\mathds{I}+\frac{\alpha_s}{4\pi\epsilon} \mathbb{H} && \text{and} &&
\gamma=\frac{\alpha_s}{2\pi} \mathbb{H}\,.
\end{align}

The operator $\mathbb{H}$ (Hamiltonian) is given by the sum of two-particles kernels,
$\mathbb{H}_{ik}^{(2\to 2)}$
\begin{equation}
\mathbb{H}^{(N\to N) }=\sum_{i,k}^{N} \mathbb{H}_{ik}^{(2\to 2)}\,.
\end{equation}
The general structure of the kernels is
\begin{eqnarray}\label{3:2-2}
\lefteqn{\mathbb{H}^{(2\to 2)}_{12}[X^{i_1}(z_1)\otimes X^{i_2}(z_2)] \,=\,}
\nonumber\\&=&
\sum_{q}\sum_{i'_1 i'_2} [C_q]^{i_1i_2}_{i'_1i'_2}
[\mathcal{H}^{(q)}_{12} X^{i'_1}\otimes X^{i'_2}](z_1,z_2)\,.
\end{eqnarray}
Here $ [C_q]^{i_1i_2}_{j_1j_2}$ is a color tensor, $\mathcal{H}^{(q)}_{12}$ is an $SL(2,\mathbb{R})$
invariant operator which acts on  coordinates of the fields, and $q$ enumerates
different structures.

Explicit expressions for the two-particle (BFLK) kernels in the light-cone gauge in coordinate space
are given in Table~\ref{table1}~\cite{Braun:2009vc}. We tacitly assume existence of $n_f$ quark flavors;
the kernels {\bf B}$^{NS}$ and  {\bf B}$^{S}$ correspond to the flavor-nonsinglet and flavor-singlet
quark pairs, respectively. In the kernel {\bf B}$^{S}$ the generators $t^a$ should be
taken $t^a=T^a$ for the $\psi\otimes \bar\psi$ pair, and $t^a=(-T^a)^t$ for the $\chi\otimes \bar\chi$
pair.


\begin{table*}[t]
\renewcommand{\arraystretch}{1.6}
\begin{tabular}{|l|c|l|}
\hline\hline
 & $X_1(z_1)\otimes X_2(z_2)$ & $\hspace*{4cm}\mathbb{H}\,[X_1\otimes X_2]\,$\\
\hline\hline
 \multirow{2}{*}{\bf A}
 &
$\psi_+\otimes\psi_+,~\psi_+\otimes\chi_+,~\bar\psi_+\otimes\bar\psi_+,$
 &
 \multirow{2}{*}{$-2(t^a_{i_1i'_1} t^a_{i_2i'_2})\Big[\widehat{\mathcal{H}}-2\sigma_q\Big]
 X^{i'_1}(z_1)\otimes X^{i'_2}(z_2)$}
\\
 &
 $\bar\psi_+\otimes\bar\chi_+,~\chi_+\otimes\chi_+,~\bar \chi_+\otimes\bar \chi_+$
 &
\\\hline
\multirow{2}{*}{{\bf B}$^{NS}$}
&
$\psi_+\otimes\bar\chi_+,\bar\psi_+ \otimes\chi_+,$
&
\multirow{2}{*}{
$-2(t^a_{i_1i'_1} t^a_{i_2i'_2}) \Big[\widehat{\mathcal{H}}-\mathcal{H}^+-2\sigma_q\Big]
 X^{i'_1}(z_1)\otimes X^{i'_2}(z_2)$}
\\
&
$\psi_+\otimes\bar\psi_+,\bar\chi_+ \otimes\chi_+$
&
\\\hline
 \multirow{3}{*}{{\bf B}$^{S}$}
&
 \multirow{3}{*}{$\psi_+\otimes\bar\psi_+,\,\chi_+\otimes\bar\chi_+$}
&
$-2(t^a_{i_1i'_1} t^a_{i_2i'_2}) \Big[\widehat{\mathcal{H}}-\mathcal{H}^+-2\sigma_q\Big]
 X^{i'_1}(z_1)\otimes X^{i'_2}(z_2)$
\\
&& $-4 t^a_{ij}\, \mathcal{H}^d\, {J}^a(z_1,z_2) $
\\
&&
$-2iz_{12} \Big\{(t^at^b)_{ij}\Big[\mathcal{H}^++\widetilde{\mathcal{H}}^+\Big]
 +2(t^bt^a)_{ij} \mathcal{H}^-\Big\}\, f^{a}_{++}(z_1)\otimes \bar f^b_{++}(z_2)$
\\\hline
\multirow{2}{*}{{\bf C}}
&
$ f^a_{++}\otimes\psi_+,\,f^a_{++}\otimes\chi_+,$
&
$-2(t^b_{aa'}\otimes t^b_{ii'})
\left[\widehat{\mathcal{H}}-\sigma_q-\sigma_g\right]X^{a'}(z_1)\otimes X^{i'}(z_2)$
\\
&
$\bar f^a_{++}\otimes \bar\psi_+,\,\bar f^a_{++}\otimes \bar\chi_+ $
&
$ - 2(t^{a'}t^a)_{ii'}P_{12}\mathcal{H}^{e,(1)}_{12}\,X^{a'}(z_1)\otimes X^{i'}(z_2)$
\\\hline
\multirow{2}{*}{{\bf D}}
&
$f^a_{++}\otimes\bar\psi_+,\,f^a_{++}\otimes\bar \chi_+,$
&
$-2(t^b_{aa'}\otimes t^b_{ii'})
\left[\widehat{\mathcal{H}}-2\mathcal{H}^+-\sigma_{q}-\sigma_g\right]X^{a'}(z_1)\otimes X^{i'}(z_2)$
\\
&
$ \bar f^a_{++}\otimes \psi_+,\,\bar f^a_{++}\otimes \chi_+ $
&
$+ 4(t^{a'}t^a)_{ii'}\mathcal{H}^{-}\,X^{a'}(z_1)\otimes X^{i'}(z_2)$
\\\hline
 {\bf E}
 &
$f^{a}_{++}\otimes f^{c}_{++},\,\bar f^{a}_{++}\otimes \bar f^{c}_{++}$
 &
 $ -2(t^b_{aa'} t^b_{cc'})
\left[\widehat{\mathcal{H}}-2\sigma_g\right] X^{a'}(z_1)\otimes X^{c'}(z_2) $
\\\hline
\multirow{3}{*}{{\bf F}}
&
\multirow{3}{*}{$f^a_{++}\otimes \bar f^c_{++}$}
&
$-2(t^b_{aa'} t^b_{cc'})\Big[\widehat{\mathcal{H}}-4\mathcal{H}^+ - 2\, \widetilde{\mathcal{H}}^+ -
2\sigma_g \Big]\,f_{++}^{a'}(z_1)\otimes \bar f_{++}^{c'} (z_2)$
\\
&&
$+12 (t^b_{ac'}t^b_{ca'}) \mathcal{H}^{-} f^{a'}_{++}(z_1)\otimes \bar f^{c'}_{++}(z_2)$
\\
&&
$+\frac{2i}{z_{12}}\Big[2\mathcal{H}^{+}P_{12}-P_{ac}\Big]
\Big(1-6\,\mathcal{H}^d\Big)\,{J}^{ac}(z_2,z_1)$
\\\hline\hline
\end{tabular}
\caption{\small Summary of the BFLK kernels \cite{Braun:2009vc}.}
\label{table1}
\renewcommand{\arraystretch}{1.0}
\end{table*}


The kernels in Table~\ref{table1} are written in terms of several
``standard'' $SL(2)$-invariant operators defined as
\begin{align}
{}[\widehat{\mathcal{H}}^{\phantom{v}}\varphi](z_1,z_2)&=
\int_0^1\frac{d\alpha}{\alpha}\Big[2\varphi(z_1,z_2)
\nonumber\\&
-\bar\alpha^{2j_1-1}\varphi(z_{12}^\alpha,z_2)-\bar\alpha^{2j_2-1}\varphi(z_1,z_{21}^\alpha)\Big]\,,
\nonumber\\
{}[{\mathcal{H}}^d\varphi](z_1,z_2)&=\int_0^1d\alpha\,
\bar\alpha^{2j_1-1}\alpha^{2j_2-1}\,\varphi(z_{12}^\alpha,z_{12}^\alpha)\,,
\nonumber\\
{}[\mathcal{H}^+\varphi](z_1,z_2)&=\int_0^1\!\!d\alpha\!\!\int_0^{\bar\alpha}\!\!d\beta\,
\bar\alpha^{2j_1-2}\bar\beta^{2j_2-2}\,\varphi(z_{12}^\alpha,z_{21}^\beta)\,,
\nonumber\\
{}[\widetilde{\mathcal{H}}^+\varphi](z_1,z_2)&=\int_0^1\!d\alpha\int_0^{\bar\alpha}\!\!d\beta\,
\bar\alpha^{2j_1-2}\bar\beta^{2j_2-2}\,\left(\frac{\alpha\beta}{\bar\alpha\bar\beta}\right)
\nonumber\\
&\phantom{=}{}\times
\varphi(z_{12}^\alpha,z_{21}^\beta)\,,
\nonumber\\
{}[\mathcal{H}^-\varphi](z_1,z_2)&=\int_0^1\!\!d\alpha\!\!\int_{\bar\alpha}^1\!\!\!d\beta\,
\bar\alpha^{2j_1-2}\bar\beta^{2j_2-2}\,\varphi(z_{12}^\alpha,z_{21}^\beta)\,,
\nonumber\\
{}[\mathcal{H}^{e,(k)}_{12}\!\varphi](z_1,z_2)&=\int_0^1d\alpha\,\bar\alpha^{2j_1-k-1}\,
\alpha^{k-1} \varphi(z_{12}^\alpha,z_2),
\nonumber\\
\label{4:HH}
\end{align}
where in the last line it is assumed that $ 0 < k < 2 j_1$. Here and below we
use a shorthand notation
$z_{ik}^\alpha=z_i\bar\alpha+z_k\alpha\,, \bar\alpha = 1-\alpha,\, z_{12}=z_1-z_2$.

The kernels depend on the conformal spins of the partons they are acting on,
$j=1$ for quarks and antiquarks and $j=3/2$ for gluons.
Another shorthand notation is
\begin{eqnarray}
\lefteqn{{J}^a(z_1,z_2)=}
\nonumber\\&=&
\sum_{A}\Big(\bar \psi^A_+(z_1) T^a\psi^A_+(z_2)+\chi^A_+(z_1) T^a\bar\chi^A_+(z_2)\Big),
\nonumber\\
\lefteqn{{J}^{ac}(z_1,z_2)=}
\nonumber\\
&=&\sum_{A}\Big[\bar\psi^A_+(z_1) T^a T^c\psi^A_+(z_2)-
\chi^A_+(z_2) T^cT^a\bar\chi^A_+(z_1)\Big],
\nonumber\\
\end{eqnarray}
where $A$ is the flavor index; the sum runs over all possible flavors.
The constants
\begin{equation}
  \sigma_q \,=\, \frac34\,,\qquad  \sigma_g \,=\, b_0/4N_c\,,
\end{equation}
where $b_0\,=\,\frac{11}{3}N_c-\frac{2}{3}n_f$, correspond to the "plus" quark field
and transverse gluon field renormalization in the axial gauge
\begin{align}
Z_q=1+\frac{\alpha_s}{2\pi\epsilon}\sigma_q C_F\,, &&
Z_g=1+\frac{\alpha_s}{2\pi\epsilon}\sigma_g C_A\,.
\end{align}
{}Finally,
$P_{12}$ and $P_{ac}$ stand for permutation operators
in position and color space, respectively. For example
\begin{eqnarray}
 P_{12}\,J^{ac}(z_1,z_2) &=& J^{ac}(z_2,z_1)\,,
\nonumber\\
 P_{ac}\,J^{ac}(z_1,z_2)&=& J^{ca}(z_1,z_2)\,.
\end{eqnarray}

Going over from the light-ray operator renormalization to
evolution equations for parton distributions
corresponds to the Fourier transformation of the functions
$\mathcal{H}^d,\mathcal{H}^{e,(k)},\widehat{\mathcal{H}},\mathcal{H}^+,\widetilde{\mathcal{H}}^+,
\mathcal{H}^-$ from coordinate to the momentum fraction representation.

A generic two-particle kernel in momentum space  has the form
\begin{align}\label{Mrep}
[\mathcal{H}\varphi](u_1,u_2)&=\int_{-\infty}^\infty \mathcal{D}v\,
\mathcal{H}(u_1,u_2|v_1,v_2)\,\varphi(v_1,v_2)\,,
\end{align}
where $\mathcal{D}v=dv_1dv_2 \delta(u_1+u_2-v_1-v_2)$.
In this Appendix we use the letters $u,v$ for momentum fractions.
Note that the integration regions over $v_1$ and $v_2$
in Eq.~(\ref{Mrep}) are (formally) infinite;
in practice they are constrained by the support properties of
parton distributions and also (Heaviside) step-functions
that are present in the kernels.
For later convenience we introduce a notation
\begin{align}\label{multitheta}
\Theta(a_1,\ldots,a_n)=\prod_{k=1}^n \theta(a_i)-\prod_{k=1}^n \theta(-a_i)\,.
\end{align}
We remind that the kernels depend on the conformal spins of fields
they acts on, cf.~Eq.~(\ref{4:HH}), so that, in principle, each one has to carry a
pair of indices $(j_1,j_2)$ which we omitted for brevity.

We obtain
\begin{equation}\label{A:Hd}
{\mathcal{H}}_{12}^d(\boldsymbol{u}|\boldsymbol{v})=
\frac{\Theta(u_1,u_2)}{u_1+u_2}\left(\frac{u_1}{u_1+u_2}\right)^{2j_1-1}
\!\!\!\left(\frac{u_2}{u_1+u_2}\right)^{2j_2-1}\!,
\end{equation}
\begin{equation}\label{A:He}
\mathcal{H}^{e,(k)}_{12}(\boldsymbol{u}|\boldsymbol{v})=\frac{\Theta(u_1,v_1-u_1)}{u_1}
\left(\frac{u_1}{v_1}\right)^{2j_1-k} \left(1-\frac{u_1}{v_1}\right)^{k-1}\!,
\end{equation}
\begin{eqnarray}\label{A:Hhat}
\lefteqn{[\widehat{\mathcal{H}}_{12}\varphi](\boldsymbol{u})=}
\notag\\
&=&\int \mathcal{D}{v}\Biggl\{
\frac{u_1}{v_1}\frac{\Theta(u_1,v_1-u_1)}{v_1-u_1}
\left(\varphi(\boldsymbol{u})-
\left(\frac{u_1}{v_1}\right)^{2j_1-2}\hspace*{-5mm}
\varphi(\boldsymbol{v})\right)
\notag\\
&+&
\frac{u_2}{v_2}\frac{\Theta(u_2,v_2-u_2)}{v_2-u_2}
\left(\varphi(\boldsymbol{u})-
\left(\frac{u_2}{v_2}\right)^{2j_2-2}\hspace*{-5mm}
\varphi(\boldsymbol{v})\right)\Biggr\}\,.
\end{eqnarray}
Note that ${\mathcal{H}}_{12}^d$ (\ref{A:Hd}) vanishes in the DGLAP region.

{}For the remaining kernels the expressions for arbitrary
conformal spins become too lengthy, so we specialize to the
particular cases of interest.

The operator $\mathcal{H}^+$ enters the BFLK kernels in Table~\ref{table1}
with the values of conformal spins $(j_1,j_2)=\{(1,1),(3/2,1),(1,3/2), (3/2,3/2)\}$.
In momentum fraction space one obtains
\begin{align}\label{A:H+}
\mathcal{H}^{+,(j_1,j_2)}(\boldsymbol{u}|\boldsymbol{v})&=
\Theta(-u_1,u_2,u_1-v_1)A^{(j_1,j_2)}(\boldsymbol{u}|\boldsymbol{v})
\notag\\
&\phantom{=}+{\Theta}(u_1,u_2,v_2-u_2)B^{(j_1,j_2)}(\boldsymbol{u}|\boldsymbol{v})
\notag\\
&\phantom{=}+{\Theta}(u_1,u_2,v_1-u_1)C^{(j_1,j_2)}(\boldsymbol{u}|\boldsymbol{v})\,.
\end{align}
The first term contributes to the DGLAP region only,
the other two --- to the ERBL region.

One derives the following expressions:
\begin{align}\label{A:H+ABC}
A^{(1,1)}(\boldsymbol{u}|\boldsymbol{v})&=\frac{v_1-u_1}{v_1v_2},
\notag\\
B^{(1,1)}(\boldsymbol{u}|\boldsymbol{v})&=\frac{u_2}{v_2(u_1+u_2)},
\notag\\
C^{(1,1)}(\boldsymbol{u}|\boldsymbol{v})&=\frac{u_1}{v_1(u_1+u_2)},
\notag\\
A^{(\frac32,1)}(\boldsymbol{u}|\boldsymbol{v})&=\frac12\frac{v_1^2-u_1^2}{v_1^2v_2},
\notag\\
B^{(\frac32,1)}(\boldsymbol{u}|\boldsymbol{v})&=\frac12\frac{u_2}{v_2}\frac{u_2+2u_1}{(u_1+u_2)^2},
\notag\\
C^{(\frac32,1)}(\boldsymbol{u}|\boldsymbol{v})&=
\frac12\frac{u_1^2}{v_1^2}\frac{v_2+2v_1}{(u_1+u_2)^2},
\notag\\
A^{(\frac32,\frac32)}(\boldsymbol{u}|\boldsymbol{v})&=\frac12\frac{v_1-u_1}{v_1^2v_2^2}
\left(v_1u_2+u_1v_2-\frac13(v_1-u_1)^2\right),
\notag\\
B^{(\frac32,\frac32)}(\boldsymbol{u}|\boldsymbol{v})&=\frac12
\left(\frac{u_2}{v_2}\right)^2\frac1{u_1+u_2}\left\{1+\frac{u_1v_2-u_2v_1/3}{(u_1+u_2)^2}\right\},
\notag\\
C^{(\frac32,\frac32)}(\boldsymbol{u}|\boldsymbol{v})&=\frac12
\left(\frac{u_1}{v_1}\right)^2\frac1{u_1+u_2}\left\{1+\frac{u_2v_1-u_1v_2/3}{(u_1+u_2)^2}\right\}.
\end{align}
Obviously
$$\mathcal{H}^{+,(1,3/2)}(u_1,u_2|v_1,v_2)=\mathcal{H}^{+,(3/2,1)}(u_2,u_1|v_2,v_1).$$

The ``modified'' $\widetilde{\mathcal{H}}^+$ kernel is needed for the conformal
spins $(j_1,j_2)=(3/2,3/2)$ only. It is given by a similar expression
\begin{align}\label{A:tildeH+}
\widetilde{\mathcal{H}}^{+,(j_1,j_2)}(\boldsymbol{u}|\boldsymbol{v})&=
\Theta(-u_1,u_2,u_1-v_1)\widetilde{A}^{(j_1,j_2)}(\boldsymbol{u}|\boldsymbol{v})
\notag\\
&\phantom{=}+{\Theta}(u_1,u_2,v_2-u_2)\widetilde{B}^{(j_1,j_2)}(\boldsymbol{u}|\boldsymbol{v})
\notag\\
&\phantom{=}+{\Theta}(u_1,u_2,v_1-u_1)\widetilde{C}^{(j_1,j_2)}(\boldsymbol{u}|\boldsymbol{v})\,
\end{align}
with
\begin{align}\label{A:tildeH+ABC}
\widetilde{A}(\boldsymbol{u}|\boldsymbol{v})&=\frac1{6}
\frac{(u_1-v_1)^3}{v_1^2v_2^2},
\notag\\
\widetilde{B}(\boldsymbol{u}|\boldsymbol{v})&=\frac12
\left(\frac{u_2}{v_2}\right)^2\frac{u_1v_2-v_1u_2/3}{(u_1+u_2)^3},
\notag\\
\widetilde{C}(\boldsymbol{u}|\boldsymbol{v})&=\frac12
\left(\frac{u_1}{v_1}\right)^2\frac{u_2v_1-v_2u_1/3}{(u_1+u_2)^3}.
\end{align}
Note that the factors $A,B,C$ (and, similar, $\widetilde{A},\widetilde{B},\widetilde{C}$)
satisfy the relation
$$
A^{(j_1,j_2)}-B^{(j_1,j_2)}+C^{(j_1,j_2)}=0\,.
$$
The kernels $\mathcal{H}^+(\boldsymbol{u}|\boldsymbol{v})$ and
 $\widetilde{\mathcal{H}}^+(\boldsymbol{u}|\boldsymbol{v})$
are continuous functions in the whole domain although they are given
by different expressions in the DGLAP and ERBL regions.

The kernel $\mathcal{H}^-$ has a different structure of regions, namely
\begin{align}\label{A:H-}
\mathcal{H}^-(\boldsymbol{u}|\boldsymbol{v})&=
\Theta(u_1,-u_2,u_2-v_1)D(\boldsymbol{u}|\boldsymbol{v})
\notag\\
&\phantom{=}+{\Theta}(u_1,u_2,v_2-u_1)E(\boldsymbol{u}|\boldsymbol{v})
\notag\\
&\phantom{=}+{\Theta}(u_1,u_2,u_1-v_2)F(\boldsymbol{u}|\boldsymbol{v})\,.
\end{align}
For the spins of interest $(j_1,j_2)=\{(3/2,1),(3/2,3/2)\}$ one gets
\begin{align}\label{A:H-DEF}
D^{(\frac32,1)}(\boldsymbol{u}|\boldsymbol{v})&=
\frac1{2v_2}\left(\frac{u_2-v_1}{v_1}\right)^2,
\nonumber\\
E^{(\frac32,1)}(\boldsymbol{u}|\boldsymbol{v})&=
\frac1{2v_2}\left(\frac{u_1}{u_1+u_2}\right)^2,
\nonumber\\
F^{(\frac32,1)}(\boldsymbol{u}|\boldsymbol{v})&=
\frac{u_2}{2v_1^2}\frac{2u_1 v_1-u_2v_2}{(u_1+u_2)^2},
\nonumber\\
D^{(\frac32,\frac32)}(\boldsymbol{u}|\boldsymbol{v})&=
\frac1{6}
\frac{(u_2-v_1)^3}{v_1^2v_2^2},
\nonumber\\
E^{(\frac32,\frac32)}(\boldsymbol{u}|\boldsymbol{v})&=
\frac12\left(\frac{u_1}{v_2}\right)^2\frac{u_2v_2-v_1u_1/3}{(u_1+u_2)^3},
\nonumber\\
F^{(\frac32,\frac32)}(\boldsymbol{u}|\boldsymbol{v})&=
\frac12\left(\frac{u_2}{v_1}\right)^2\frac{u_1v_1-v_2u_2/3}{(u_1+u_2)^3}.
\end{align}

We also give here the explicit expressions for the off-diagonal kernels~(\ref{Off}) in
momentum representation. The $\mathbb{H}_{QF}^\pm$--kernel can be written as
\begin{align}\label{A:QF}
\mathbb{H}_{QF}^\pm(\boldsymbol{u}|\boldsymbol{v})&=n_f
\left(\mathcal{V}^+_{13}(\boldsymbol{u}|\boldsymbol{v})\mp
\mathcal{V}^-_{13}(\boldsymbol{u}|\boldsymbol{v})\right)\,,
\end{align}
where $\mathcal{V}^+_{13}(\boldsymbol{u}|\boldsymbol{v})$ and
$\mathcal{V}^-_{13}(\boldsymbol{u}|\boldsymbol{v})$ have the
decomposition over different regions of the form~(\ref{A:H+}) and
(\ref{A:H-}), respectively. For the corresponding functions $A_{\mathcal{V}},B_{\mathcal{V}},C_{\mathcal{V}}$ and
$D_{\mathcal{V}},E_{\mathcal{V}},F_{\mathcal{V}}$ one obtains
\begin{align}\label{}
{A}_{\mathcal{V}}(\boldsymbol{u}|\boldsymbol{v})&=\phantom{-}\frac{u_1 u_3}{v_1^2 v_3^2}\,,
\nonumber\\
{B}_{\mathcal{V}}(\boldsymbol{u}|\boldsymbol{v})&=\phantom{-}
\frac{u_1 u_3}{v_3^2}\frac{v_1+3v_3}{(u_1+u_3)^3}\,,
\nonumber\\
{C}_{\mathcal{V}}(\boldsymbol{u}|\boldsymbol{v})&=-
\frac{u_1 u_3}{v_1^2}\frac{v_3+3v_1}{(u_1+u_3)^3}\,,
\nonumber\\	
{D}_{\mathcal{V}}(\boldsymbol{u}|\boldsymbol{v})&=\phantom{-}\frac{(u_3-v_1)^2}{v_1^2v_3^2}\,,
\nonumber\\
{E}_{\mathcal{V}}(\boldsymbol{u}|\boldsymbol{v})&=-
\frac{u_1^2}{v_3^2(u_1+u_3)^2}\left[\frac{2u_3v_3}{u_1(u_1+u_3)}-1\right],
\nonumber\\
{F}_{\mathcal{V}}(\boldsymbol{u}|\boldsymbol{v})&=\phantom{-}
\frac{u_3^2}{v_1^2(u_1+u_3)^2}\left[\frac{2u_1v_1}{u_3(u_1+u_3)}-1\right].
\end{align}
The $\mathbb{H}_{FQ}^\pm$--kernel can be written as
\begin{align}\label{HFQ}
\mathbb{H}_{FQ}^+&=N_c(1-P_{23})\left[\mathcal{W}^++\mathcal{W}^--2\Delta\mathcal{W}\right],
\nonumber\\
\mathbb{H}_{FQ}^-&=-\frac{N_c^2-4}{N_c}(1+P_{23})\left[\mathcal{W}^++\mathcal{W}^-\right].
\end{align}
Here
\begin{align}\label{}
\Delta\mathcal{W}(\boldsymbol{u}|\boldsymbol{v})&=\Theta(u_3,v_3-u_3)
-\Theta(u_1,u_3)\frac{u_1^2(3u_3+u_1)}{(u_1+u_3)^3}\,.
\end{align}
The kernel $\mathcal{W}^+(u_1,u_3|v_1,v_3)$ has the structure~(\ref{A:H+})
with
\begin{align}\label{}
A_{\mathcal{W}}(\boldsymbol{u}|\boldsymbol{v})=1\,,&&
B_{\mathcal{W}}(\boldsymbol{u}|\boldsymbol{v})=-C_{\mathcal{W}}(\boldsymbol{u}|\boldsymbol{v})=\frac12\,,
\end{align}
and the kernel $\mathcal{W}^-(u_1,u_3|v_1,v_3)$ takes the form~(\ref{A:H-}) with
\begin{align}\label{A:last}
D_{\mathcal{W}}(\boldsymbol{u}|\boldsymbol{v})&=-\frac{(v_1-u_3)^2}{v_1 v_3}\,,
\nonumber\\
E_{\mathcal{W}}(\boldsymbol{u}|\boldsymbol{v})&=\phantom{-}\frac12-\frac{u_1^2}{v_3(u_1+u_3)}\,,
\nonumber\\
F_{\mathcal{W}}(\boldsymbol{u}|\boldsymbol{v})&=-\frac12+\frac{u_3^2}{v_1(u_1+u_3)}\,.
\end{align}
The kernels $\mathcal{W}^\pm, \Delta\mathcal{W}$ are antisymmetric under
$(u_1,v_1)\leftrightarrow (u_3,v_3)$. Note that the kernels in~(\ref{HFQ}) have
discontinuities on the DGLAP--ERBL boundaries.

\noindent


\end{document}